\documentclass{article}
\usepackage[]{amsmath,amssymb}
\usepackage{amsfonts}
\usepackage{graphics,epsfig}
\def\ie{{\rm i.e.,\/}\ }
\def\etc{{\rm etc.\/}\ }

\newcommand{\qch}{\check{q}}
\newcommand{\hch}{\check{h}}
\def \otimesdot {\stackrel{\cdot}{\otimes}}
\newcommand{\ZZ}{\mathbb{Z}}

\newcommand{\CC}{\mathbb{C}}

\begin{document}
\pagestyle{empty}
\begin{flushright}
{CERN-TH/2000-179}\\
{CPT-2000/P.4077}\\
math-ph/0011006\\
\end{flushright}
\vspace*{5mm}
\begin{center}
{\bf NOTES ON THE QUANTUM TETRAHEDRON} \\
\vspace*{0.2cm} 
{\bf R. Coquereaux}$^{*}$ \\
\vspace{0.3cm}
Theoretical Physics Division, CERN \\
CH - 1211 Geneva 23 \\
and\\
Centre de Physique Th\'eorique,CNRS, Luminy, Case 907 \\
       F-13288 Marseille Cedex 9,France\\ 
\vspace*{1.3cm}  
{\bf ABSTRACT} \\ \end{center}
\vspace*{5mm}
\noindent
This is a set of notes describing several aspects of the space of 
paths on ADE Dynkin diagrams, with a particular attention 
paid to the graph $E_6$. Many results originally due to A. Ocneanu 
are here described in a very elementary way (manipulation
of square or rectangular matrices). We recall the concept of 
essential matrices (intertwiners) for a graph and describe their module properties
with respect to right and left actions of fusion algebras.
In the case of the graph $E_6$, essential matrices build up a right 
module
with respect to its own fusion algebra but a left module with respect to 
the fusion algebra of $A_{11}$.
 We  present two original results: 
 1) Our first contribution is to show how to recover the 
Ocneanu graph of quantum symmetries 
of the Dynkin diagram $E_6$ from the natural multiplication defined 
in the tensor square of its fusion algebra (the
tensor product should be taken over a particular subalgebra); 
 this is the Cayley graph for the two generators of
the twelve dimensional algebra $E_6 \otimes_{A_3} E_6$ (here $A_3$ 
and $E_6$ refer to the commutative
fusion algebras of the corresponding graphs). 
2) To every point of the graph of quantum symmetries one can 
associate a particular  matrix describing the `` torus structure'' of the 
chosen Dynkin diagram; following Ocneanu, one obtains in this way,
in the case of $E_6$, 
twelve such matrices of dimension $11\times 11$, one of them is a 
modular invariant and encodes the partition function of
the corresponding conformal field theory. 
Our own next contribution is to provide a simple algorithm for the 
determination of these matrices.

\vspace*{0.9cm}

\vfill\eject

\setcounter{page}{1}
\pagestyle{plain}

\section{Introduction}

 \subsection{Summary} 
 \label{sec:summary}
One purpose of the present paper is to present a simple construction 
for the ``Ocneanu graph''
describing the quantum symmetries of the Dynkin diagram $E_6$.
Another purpose of our paper is to give a simple method allowing one to determine,
for each point of this Ocneanu diagram, a particular ``toric 
matrix''. One of these matrices, associated
with the origin of the graph, is a modular invariant. 
These toric matrices were first obtained by A. Ocneanu (they were 
never made available in printed form).
Our techniques bring some simplification to the calculations 
and should improve the understanding of the inter-related structures
appearing in this subject.
We choose to follow the example of $E_6$ because it
exhibits quite generic features\footnote{Note added in proof: 
the calculations and results, for all ADE Dynkin diagrams, can be 
found in our subsequent paper \cite{Coque-Gil:ADE}}. 

We do not intend to give here a detailed account of
the constructions of Ocneanu but the present paper may provide a simple 
introduction to this theory since it does not
require any particular knowledge of operator algebra or conformal
field theory. The mathematical background needed  here  usually does 
not involve more than multiplication of rectangular matrices.
Therefore, apart from the two above mentioned results, another purpose 
of our paper is to show how one can recover several important results
of this theory, bypassing many of the steps described in reference 
~\cite{Ocneanu:paths}.

\subsection{Historical comments}

The classification of conformal field theories of $SU(2)$ type was 
performed by A.Cappelli, C.Itzykson and J.B. Zuber in ~\cite{CIZ}.
They established a correspondance between  these conformal field 
theories and the ADE Dynkin diagrams
used in the classification of simple Lie algebras. This correspondance
was found, at the same time, by V. Pasquier in ~\cite{Pasquier}, 
within the context of lattice gauge models. 
This new ADE classification has then been discussed in
several papers (see in particular \cite{Slodowy}).
Later, a generalization of  lattice models to the case of $SU(n)$ was 
studied in \cite{PasquierSUn}
and  the study of conformal theories of $SU(3)$ type was 
performed by P. Di Francesco and J.B. Zuber in ~\cite{DiFrancescoZuber}
 (see also ~\cite{Zuber:Bariloche} and reference therein); the last 
authors  related 
the $SU(3)$ classification problem to new types of graphs 
generalizing the $ADE$ Dynkin diagrams. 
Several years ago, in order to study Von Neumann algebras, a theory 
of ``paragroups'' was 
invented by A. Ocneanu ~\cite{Ocneanu:paragroups}.  Roughly speaking 
these paragroups characterize embeddings of operator algebras.
The combinatorial data provided by Dynkin diagrams (and 
corresponding affine Dynkin diagrams) provides a simple example 
of this general framework. Many details
have been worked out by A. Ocneanu himself who gave several talks on 
the  subject (for instance ~\cite{Ocneanu:Marseille})
but this work was not made available in written form,
with the exception of a recent set of notes ~\cite{Ocneanu:paths}.
Later,  A. Ocneanu discovered how to
recover  the ADE classification of modular invariant partition functions 
 from his theory of ``Quantum Symmetries'' on 
graphs and, more recently ~\cite{Ocneanu:Bariloche},
how to generalize his method to conformal theories of type $SU(3)$ 
and $SU(4)$, therefore establishing a direct relation with the results 
of~\cite{DiFrancescoZuber}.

 \subsection{Structure of the paper} 

In the first part we show how to construct a particular finite 
dimensional 
commutative algebra (technically an hypergroup) 
from the combinatorial data provided by the graph $E_6$; this algebra 
can be realized in terms of a $6\times 6$ 
commuting matrices that will be called ``graph fusion matrices'' (we shall 
recover, in the corresponding
subsection, several results that are more or less well known, and
they can be found in
the book ~\cite{FMS:book}). These matrices can be used to study
paths on the $E_6$ graph.
In the second part we use the concept of essential paths (due to 
Ocneanu) to define  what we  call ``essential matrices'' ($E_a$): 
These
are rectangular $11 \times 6$ matrices that generate a bimodule with 
respect to the fusion algebra of the $E_6$ graph
(from one side) and with respect to the fusion algebra of the 
$A_{11}$ graph (from the other side).
In the third part we build the commutative algebra $E_6 \otimes_{A_3} 
E_6$ where $E_6$ refers to the fusion algebra of the
$E_6$ graph and $A_3$ refers to a particular subalgebra (isomorphic 
with the fusion 
algebra of the $A_3$ graph); the tensor product
is taken above $A_3$ so that this algebra has dimension $6 \times 6 
/3 = 12$. Its multiplicative structure is described by a graph with 
$12$ points describing the quantum symmetries of the $E_6$ graph; 
this graph 
was originally obtained by A. Ocneanu 
after diagonalization of the convolution product in 
the bigebra of endomorphisms of essential 
paths (a kind
of generalized finite dimensional
Racah-Wigner bigebra of dimension $2512$).
Our approach based on the study 
of the finite dimensional commutative
algebra  $E_6 \otimes_{A_3} E_6$ allows one to obtain directly the 
Ocneanu graph, therefore bypassing the
rather complicated study of the Racah-Wigner bigebra (one of the two
multiplications of the later 
involves generalized $6j$ symbols containing $24$-th roots of unity).
We also give an interpretation for the square matrices  
(respectively of size $(11,11)$ or $(6,6)$) obtained when calculating products
of essential matrices $E_a. \widetilde E_b$ or 
$\widetilde E_a.E_b$; the symbol $\widetilde {}$ stands for ``transpose'' (we 
sometimes use the symbol ${}^{T}$). 
Actually, tables $E\times E \rightarrow A$ and $E \times E \rightarrow S$ given 
in section \ref{sec: EEA} and \ref{sec: EES} describe two different sets of 
vertices ((\ref{fig:EEA}) or (\ref{fig:EES})),
out of which one can nicely encode all the  elements of the two 
different adapted basis for the above bigebra (sometimes called the ``double 
triangle algebra'').

In the fourth part we define ``reduced essential matrices'' by
removing from the  essential matrices of $E_6$ the columns associated 
to the
supplement of its $A_3$ subalgebra 
and use them to construct twelve ``toric matrices'' $11\times 
11$ (one for each point of the Ocneanu graph). 
One of these matrices is a modular invariant, in the sense that it 
commutes with the generators $S$ and $T$ of $SL(2,\ZZ)$, in the 
$11$-dimensional representation of Hurwitz-Verlinde.
This particular matrix is associated with the unit of the  $E_6 
\otimes_{A_3} E_6$ algebra and defines a
modular invariant sesquilinear form wich is nothing else than the 
partition function of Cappelli, Itzykson, Zuber.
The other toric  matrices (associated with the other points of the 
Ocneanu graph) are also very interesting (see footnote $3$),  
but are not invariant under $SL(2,\ZZ)$.
In the last section (Comments), we gather miscellaneous comments 
about the relation between our approach and
the one based on the study of the generalized Ocneanu-Racah-Wigner bigebra.
We conclude with several open questions concerning
an interpretation in terms of non 
semi-simple (but finite dimensional) quantum groups.

\subsection{Remarks}
 \label{sec:remarks}
The reader is probably aware of the fact that the seven points of the 
affine graph $E_6^{(1)}$ are in one to one correspondance (McKay 
correspondance ~\cite{McKay})
with the irreducible representations
of the binary tetrahedral group (the two-fold covering of the 
tetrahedral group), and that the corresponding
fusion algebra specified by this affine Dynkin diagram encodes the 
structure of the Grothendieck ring of
representations of this finite group (of order $24$). When the 
graph $E_6$ is replaced by the graph $E_6^{(1)}$, many
constructions described in this paper can be
interpreted in terms of conventional finite group theory.
The binary tetrahedral group is a rather classical (well-known) 
object, even if its treatment based on the structure
of the affine $E_6^{(1)}$ graph is not so well known. The interested 
reader can refer 
to ~\cite{Coquereaux:ClassicalTetra} for a ``non standard'' 
discussion of the properties
of this finite group, along these lines.
Removing one node from this affine graph (hence getting $E_6$ itself) 
leads
to entirely new results and to what constitutes the subject of the 
present article.
It would be certainly useful to carry out the analysis in parallel, 
for both affine and non affine Dynkin diagrams but
this would dangerously increase the size of this paper\ldots We shall 
nevertheless make several remarks about the group
case situation, all along the text, that should help the reader to 
perform fruitful  analogies and develop some intuition.
The above remark partly justifies our title for the present paper.

Actually, an Ocneanu graph
encoding quantum symmetries usually involves ``connections'' between 
a  pair of diagrams (for 
instance, in the case 
$E_6 \otimes_{A_3} E_6$, it involves twice the graph $E_6$ itself). 
These two graphs should have same Coxeter number:
We have distinct theories for the pairs 
$A_{11}-A_{11}$, $A_{11}-E_6$, $D_5-D_5$, $A_{11}-D_5$,
$D_5-E_6$ and $E_6-E_6$. Here we only describe part of the results 
relative to the
$A_{11}-A_{11}$, $A_{11}-E_6$ and $E_6-E_6$ situations (and
especially the last one).
The reader may want to know why we restrict
 our study to these cases  and do not present  a full 
description of the situation in all $ADE$ cases. One reason is 
somehow pedagogical: we believe that it is useful
to grasp the main ideas by studying a particular
case that exhibits generic features. 
Another reason is size: corresponding 
calculations, or even presentation of results, can be rather 
long\footnote{See footnote in section \ref{sec:summary}}.
A last reason is anteriority: we believe
that many results concerning the $ADE$ (or affine $ADE$), 
have been fully worked out
by A. Ocneanu himself (certainly using other techniques) and will 
-- maybe -- appear some day.
Our modest contribution should 
allow the dedicated reader to recover many results in a simple way.

Let us mention that the study of the $E_8$ case  is very 
similar to the $E_6$ case: for $E_8$, 
 the Ocneanu graph posesses $32 = 8\times 8/2$ points, one for each 
element of $E_8 \otimes_{A_2} E_8$.
The study of $A_{N}$ is also similar but is a bit too ``simple'' since
several interesting constructions just coincide in that case.
Diagrams $E_7$ and 
$D_{odd}$ are special because their fusion algebra is
not a positive hypergroup (this was first noticed long ago, using another 
terminology by ~\cite{Pasquier}) but only a module over
an hypergroup. The case of $D_{even}$ is also special because the 
algebra associated with its Ocneanu graph is not commutative.
In order to study them, the techniques that 
we introduce for $E_{6}$ have to be slightly modified (see 
\cite{Coque-Gil:ADE}).

One interesting direction of research is to generalize the simple 
algorithms developed here to recover and generalize the results
relative to conformal field theories with chiral algebra
 $SU(3)$, $SU(4)$, \ldots 
(see ~\cite{DiFrancescoZuber}, ~\cite{Ocneanu:Bariloche},
~\cite{PetkovaZuber} and the lectures of  J.B. Zuber at Bariloche
~\cite{Zuber:Bariloche})\footnote{After completion of 
the present work,
we received a new preprint
by V.B. Petkova and J.B. Zuber (\cite{PetkovaZuber:Oc}), 
giving a physical interpretation of these
other toric matrices in terms of
partition functions associated with twisted boundary conditions 
(defect lines) in 
boundary conformal field theories}.
Details concerning $SU(n)$ generalizations, 
following the methods explained in the present paper,
should appear in ~\cite{Schieber:thesis}.

\section{The graph $E_6$ and its fusion algebra}
 
\subsection{The graph}
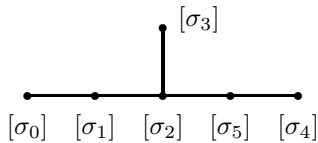
\begin{figure}[htb]
\unitlength 0.6mm
\begin{center}
\begin{picture}(95,35)
\thinlines 
\multiput(25,10)(15,0){5}{\circle*{2}}
\put(55,25){\circle*{2}}
\thicklines
\put(25,10){\line(1,0){60}}
\put(55,10){\line(0,1){15}}
\put(25,3){\makebox(0,0){$[\sigma_0]$}}
\put(40,3){\makebox(0,0){$[\sigma_1]$}}
\put(55,3){\makebox(0,0){$[\sigma_2]$}}
\put(70,3){\makebox(0,0){$[\sigma_5]$}}
\put(85,3){\makebox(0,0){$[\sigma_4]$}}
\put(63,27){\makebox(0,0){$[\sigma_3]$}}
\end{picture}
\caption{The graph of ${E_6}$}
\label{graphE6}
\end{center}
\end{figure}
The labelling of the vertices $\sigma_a$
of the graph  (Fig. \ \ref{graphE6})
follows the convention
$(0,1,2,5,4;3)$.
The reader should distinguish this labelling from the order itself 
that we have chosen to enumerate the
vertices (\ie, for instance, the fourth vertex in the list  is 
called  $\sigma_5$). To each vertex $\sigma_a$ we associate
a basis (column) vector $V_a$ in a six dimensional vector space; 
for instance $V_0= (1,0,0,0,0,0)^{T}$,\ldots,  $V_5 = 
(0,0,0,1,0,0)^{T}$, \ldots

The adjacency matrix of this graph (we use the above order for 
labelling the 
vertices) is: 
$$
G = 
 \left( \begin{array}{cccccc}
 0&1&0&0&0&0  \\
 1&0&1&0&0&0  \\
 0&1&0&1&0&1  \\
 0&0&1&0&1&0  \\
 0&0&0&1&0&0  \\
 0&0&1&0&0&0   
 \end{array} \right)
$$

\subsection{Norm of the graph and Perron Frobenius eigenvector}

The norm of this graph is, by definition, the largest eigenvalue of 
its adjacency matrix.
It is 
$$
\beta = \frac{{\sqrt 3}+1}{\sqrt 2}
$$

For a given Dynkin diagram, it is convenient to set
 	$\beta = \qch + 1/\qch$,
	with $\qch = e^{\hch}$ and $\hch = i \pi/N$.
 	Then $\beta = 2 \cos(\pi/N)$
	and $\qch$ is a root of unity. 
	 In the present case (graph $E_6$), $N=12$,
	indeed $$ 2 \cos(\pi/12) =  \frac{{\sqrt 3}+1}{\sqrt 2}$$
	 Notice that $N=12$ is the dual Coxeter number of $E_6$ (notice that 
we do not
	need to use any knowledge coming from the theory of Lie algebras).
  	In all cases (other Dynkin diagrams), $\beta$ is equal to the 
``$q$'' 
	number $[2]_{\qch}$ (where $[n]_{\qch} \doteq \frac {\qch^n - 
\qch^{-n}} {\qch - 
	\qch^{-1}}$).
	Warning: we  set $h = 2 \hch$ and $q = \qch^{2}$, so that 
	$q = e^{2i \pi/N}$ whereas $\qch = e^{i \pi/N}$.
In the case of affine Dynkin diagrams (for instance $E_6^{(1)}$), 
$\beta$ is always equal to $2$.

The Perron Frobenius eigenvector $D$
 is, by definition the corresponding normalized eigenvector
(the normalization consists in setting $D_0 = 1$.
	One finds
	$$
	D = \{1,  \frac{{\sqrt 3}+1}{\sqrt 2}, 1+\sqrt 3,  
\frac{{\sqrt 3}+1}{\sqrt 2}, 1; \sqrt 2\}
	$$
It is nice to write it in terms of $q$-numbers (with $N=12$), one 
finds\footnote{We have suppressed the sub-index
$\qch$ from the brackets}
	$$
	D = \{ [1],[2],[3],[2],[1];[3]/[2]\}
	$$
The component of $D$ associated with the origin $\sigma_0$ of the 
graph is minimal.

In the case of $E_6^{(1)}$, the seven entries of $D$ are the (usual) 
integers $1,2,3,2,1,2,1$;
 	these numbers coincide with the dimensions of
	the seven irreducible representations (``irreps'') of the binary 
tetrahedral group.
	For this reason, the six entries of $D$, in the $E_6$ case, should 
be thought of as 
	quantum dimensions for the irreps of  a quantum analogue of this 
finite group.
	
Returning to the $E_6$ case, we notice that the  eigenvalues of the 
adjacency matrix $G$ read
$2 \cos \pi \frac{m}{N}$, with $N=12$ and $m=1,4,5,7,8,11$. The value 
obtained with
 $m=1$ gives the norm of the graph. These integers are also the 
Coxeter 
exponents of the Lie group
$E_6$.

\subsection{ Hypergroup structure}
In the classical case (\ie the group case), irreducible 
representations 
can be tensorially multiplied
and decomposed into sum of irreps; by considering sums and difference,
 	they actually generate a commutative ring (or a commutative 
algebra): the Grothendieck ring of virtual
	characters. By analogy, we  build a finite dimensional 
associative and commutative algebra
	with the above combinatorial data. This was first done, to our 
knowledge, by V. Pasquier (~\cite{Pasquier}) who
	also noticed that this construction is not always possible: it
	works for all affine Dynkin diagrams (in which case
	one recovers the multiplication of characters of the binary groups 
of Platonic bodies) and for Dynkin
	diagrams of type $A_n$, $D_{2n}$, $E_6$ and $E_8$. However, for
	$E_7$ and $D_{odd}$, it does not work 
in the same way (lack of positivity of structure constants in the 
$E_7$ case).

	We restrict ourselves to the $E_6$ case and build this algebra as 
follows (we shall call it the
	fusion algebra  associated with the graph $E_6$):
	\begin{itemize}
	\item The algebra is linearly generated by the six elements 
$\sigma_0,\ldots,\sigma_5$.
	\item $\sigma_0$ is the unit.
	\item $\sigma_1$ is the algebraic generator: multiplication by 
$\sigma_1$ is given by the adjacency matrix (this
 	is nothing else than an eigenvalue equation for $G$),
	$G.(\sigma_{0},\sigma_{1},\ldots 
	\sigma_{5})=\sigma_{1}(\sigma_{0},\sigma_{1},\ldots \sigma_{5})$.
	More explicitly:
	$ \sigma_1 \sigma_0 = \sigma_1$	, $ \sigma_1 \sigma_1 = \sigma_0 + 
\sigma_2$, $\sigma_1 \sigma_2 = \sigma_1
	+\sigma_3 +\sigma_5$,
	$ \sigma_1 \sigma_5 = \sigma_2 + \sigma_4$, $ \sigma_1 \sigma_4 = 
\sigma_5$, $ \sigma_1 \sigma_3 = \sigma_2.$	
	The reader will notice that this is nothing else than a quantum 
analogue of multiplication of spins (in the case of
	$SU(2)$, the corresponding graph is the infinite graph $A_\infty = 
\tau_0, \tau_1,\ldots$
	 where the point $\tau_{2j}$ refers to the irrep of spin $j$ (of 
dimension $2j+1$),
 and where $\tau_1 \tau_q = \tau_{q-1} +
	\tau_{q+1}$ (composition of an arbitrary spin with a spin $1/2$). 
	\item Once multiplication by the generator $\sigma_1$ is known, one 
may multiply arbitrary $\sigma_p$'s by
 	imposing associativity and commutativity of the algebra.
	For instance: 
\begin{eqnarray*}
	\sigma_2 \sigma_2 & = & (\sigma_1 \sigma_1 - \sigma_0)\sigma_2 = 
\sigma_1\sigma_1\sigma_2 - \sigma_2
	= \sigma_1(\sigma_1+\sigma_3+\sigma_5)-\sigma_2 \\
{} &=& \sigma_0+\sigma_2+\sigma_2+\sigma_2+\sigma_4-\sigma_2
	=\sigma_0 +2 \sigma_2 + \sigma_4
	\end{eqnarray*}
	\end{itemize}

 This fusion algebra (or graph algebra) is a particular 
example of what is
called a commutative positive integral hypergroup (see general 
definitions in the collection of papers
~\cite{hypergroups}); prototype of commutative
hypergroups are the class hypergroup and the representation 
hypergroup of a group (which is also
the Grothendieck ring of its virtual characters).
	
\subsection{The $E_6 \times E_6 \mapsto E_6$ multiplication table}

	In this way, one can construct the following multiplication table 
(we write $a$ rather than $\sigma_a$), that
	we call the fusion table for $E_6$:

$$
\begin{array}{||c||c|c|c||c|c|c||}
\hline
E_6 & 0 & 3  & 4 & 1 & 2 & 5   \\
\hline
\hline
0 & 0 & 3 & 4  & 1 & 2 & 5 \\
3 & 3 & 04  & 3 & 2 & 1 5 & 2 \\
4 & 4 & 3 & 0 & 5 & 2 & 1 \\
\hline
1 & 1 & 2 & 5 & 0 2 & 1 3 5 & 2 4   \\
2 & 2 & 1 5 & 2 & 1 3 5 & 0224 & 1 3 5 \\
5& 5  & 2 & 1 & 2 4 & 1 3 5 & 0 2 \\
\hline
\end{array}
$$
Notice that all entries are positive integers; this is not trivial 
(and fails to be true for graphs
	of type $E_7$ or $D_{odd}$).
The structure constants of the fusion algebra are the integers 
$C_{abc}$ that appear in the previous multiplication
table ($\sigma_a \sigma_b = C_{abc} \sigma_c$); 
we have for instance $\sigma_2 \sigma_2 = \sigma_0 + 2 \sigma_2 + 
\sigma_4$, therefore $C_{220}=1$, $C_{222}=2$,
$C_{224}=1$ and the other $C_{22c}$ are equal to zero.

	Notice also that we have chosen the order $\{0 3 4 1 2 5 \}$ to 
display this multiplication table;
	the reason is that it shows clearly that $\{0 3 4 \}$ generate a 
subalgebra with particular properties; we shall
	come back to this later.

From the above table, we can check that	
\begin{eqnarray*}
\sigma_0 &=& 1 \\
\sigma_1 &=& \sigma_1 \\
\sigma_2 &=& \sigma_1.\sigma_1 - \sigma_0 \\
\sigma_4 &=& \sigma_1.\sigma_1.\sigma_1.\sigma_1 - 4 
\sigma_1.\sigma_1 + 2 \sigma_0 \\
\sigma_5 &=& \sigma_1.\sigma_4 \\
\sigma_3 &=&  - \sigma_1.(\sigma_4 - \sigma_1.\sigma_1 + 2 \sigma_0) 
\end{eqnarray*}

\subsection{Graph fusion matrices and paths on the graph $E_6$} 

\paragraph {}
What actually turn out to be useful are the integral matrices $N_{a}$;
in the present case ($E_{6}$ case) these are symmetric 
matrices\footnote{Warning: indices do not refer to row and line numbers
but to the labels $(0,1,2,5,4,3)$ of vertices}.
$$
(N_{a})_{bc}= C_{abc}
$$
Because of the algebraic relations satisfied by the generators 
$\sigma_a$, the simplest way to obtain the six
$6 \times 6$ matrices $N_a$ is to set:
\begin{eqnarray*}
N_0 &=& Id_6 {\textstyle (the \, identity \, matrix)} \\
N_1 &=& G \\
N_2 &=& G.G - N_0 \\
N_4 &=& G.G.G.G - 4 G.G + 2 N_0 \\
N_5 &=& G.N_4 \\
N_3 &=&  - G.(N_4 - G.G + 2 N_0) 
\end{eqnarray*}
Again one should notice that $N_0, N_3$ and $N_4$ form a subalgebra 
(graph $A_3$):
\begin{eqnarray*}
N_3.N_3 &=&  N_0 + N_4\\
N_4.N_3 &=& N_3 \\
N_4.N_4 &=& N_0
\end{eqnarray*}

Using the ordered basis $(012543)$, we have
{\small
$$
\begin{array}{ccc}
N_0 = \left( \begin{array}{cccccc}
  1&0&0&0&0&0 \\
   0&1&0&0&0&0 \\
   0&0&1&0&0&0 \\
   0&0&0&1&0&0 \\
   0&0&0&0&1&0 \\
   0&0&0&0&0&1 
  \end{array} \right)
&
N_1 = \left( \begin{array}{cccccc}
  0&1&0&0&0&0\\
   1&0&1&0&0&0\\
   0&1&0&1&0&1\\
   0&0&1&0&1&0\\
   0&0&0&1&0&0\\
   0&0&1&0&0&0 
  \end{array} \right)
&
N_2 = \left( \begin{array}{cccccc}
  0&0&1&0&0&0\\
   0&1&0&1&0&1\\
   1&0&2&0&1&0\\
   0&1&0&1&0&1\\
   0&0&1&0&0&0\\
   0&1&0&1&0&0 
  \end{array} \right)
\end{array}
$$
$$
\begin{array}{ccc}
N_5 = \left( \begin{array}{cccccc}
  0&0&0&1&0&0\\
   0&0&1&0&1&0\\
   0&1&0&1&0&1\\
   1&0&1&0&0&0\\
   0&1&0&0&0&0\\
   0&0&1&0&0&0 
  \end{array} \right)
&
N_4 = \left( \begin{array}{cccccc}
  0&0&0&0&1&0\\
   0&0&0&1&0&0\\
   0&0&1&0&0&0\\
   0&1&0&0&0&0\\
   1&0&0&0&0&0\\
   0&0&0&0&0&1 
  \end{array} \right)
&
N_3 = \left( \begin{array}{cccccc}
  0&0&0&0&0&1\\
   0&0&1&0&0&0\\
   0&1&0&1&0&0\\
   0&0&1&0&0&0\\
   0&0&0&0&0&1\\
   1&0&0&0&1&0 
  \end{array} \right)
\end{array}
$$}

\paragraph{} The ring generated by matrices $N_{a}$ provides a 
faithful 
matrix realization
of the fusion algebra.
In particular, the Dynkin diagram of $E_{6}$, considered as the 
graph of multiplication by $\sigma_1$ is also the  graph of 
multiplication by the matrix $N_1$
(indeed, $N_1.N_1 = N_0 + N_2$, etc).

Warning: In this paper, the notation $E_6$ will denote  the Dynkin 
diagram $E_6$,
its fusion algebra, also called graph algebra (the commutative algebra generated by the  
$\sigma_a$),  or the explicit matrix algebra generated by the $N_a$ matrices. 
The context should be clear enough
to avoid ambiguities.

 \label{sec:E6}

Since all these $N_{a}$ matrices commute
with 
one another, they can be
simultaneously diagonalized.
If we were working with the graph $E_6^{(1)}$ rather than with $E_6$, 
\ie
in the finite group case (the binary tetrahedral group), the 
simultaneous diagonalization 
$S^{-1}.N_a.S$
of the $N_a$ matrices
would
be done thanks to a matrix $S$, which is nothing else than the
character table (this is precisely the method used to recover a
character table from the structure constants of the Grothendieck ring
when the multiplication of the group itself is not known). 
In the present case (graph of $E_6$), the $6\times 6$ matrix $S$ is a 
kind of ``non commutative'' 
character table (or Fourier transform); this matrix
is clearly an interesting  object but we shall
not make an explicit use of it in the sequel. 
The matrix $S$ associated with the graph $A_{11}$ will appear in  
section \ref{sec:A11}  and play
an important role later on (Verlinde representation of the modular 
group).

\paragraph{} A last comment about the fusion algebra of the graph 
$E_6$: it
is isomorphic with the algebra $\CC[X]/P[X]$ of complex polynomials
modulo $P[X]$ where $P[X]=(X^2-1)(X^4-4 X^2 + 1)$ is the 
characteristic
polynomial of the matrix $G$; this property (already mentionned in 
~\cite{FMS:book})
 is a direct consequence of the Cayley Hamilton theorem.

\subsection{The $A_3$ subalgebra}

We already noticed that the algebra generated by the vertices 
$\sigma_0, \sigma_3$ and $\sigma_4$
is a subalgebra of the fusion algebra of the $E_6$ graph. This 
algebra is actually isomorphic
to the fusion algebra of the Dynkin graph $A_3$. This is almost 
obvious: consider the following figure (Fig. \ \ref{graphA3}).
%
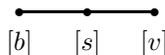
\begin{figure}[htb]
\unitlength 0.6mm
\begin{center}
\begin{picture}(95,35)
\thinlines 
\multiput(25,10)(15,0){3}{\circle*{2}}
\thicklines
\put(25,10){\line(1,0){30}}
\put(25,3){\makebox(0,0){$[b]$}}
\put(40,3){\makebox(0,0){$[s]$}}
\put(55,3){\makebox(0,0){$[v]$}}
\end{picture}
\caption{The graph of ${A_3}$}
\label{graphA3}
\end{center}
\end{figure}

The corresponding fusion algebra is therefore defined by the relations
$b s = s$, ${s}^2=b + v$, $s 
v = s$.
This implies $v v = (s s - b)v = s 
s - v = b$.
The correponding multiplication table is the same as the one obtained 
by restriction of
the $E_6$ table to the vertices $0,3,4$, under the identification 
$b \rightarrow \sigma_0$,
$s \rightarrow \sigma_3$, $v \rightarrow \sigma_4$.

The fusion subalgebra $A_3$ of the fusion algebra of the graph $E_6$ 
has also the following
remarkable property: Call $\cal P$ the vector space linearly 
generated by $\{\sigma_1,\sigma_2,\sigma_5\}$.
From the table of multiplication, we see that 
\begin{eqnarray*}
E_6 &=& A_3 \oplus {\cal P} \\
A_3 . A_3 &=& A_3 \\
A_3 . {\cal P} &=&  {\cal P} . A_3 =  {\cal P} 
\end{eqnarray*}
Such properties have been already described in various places
(see ~\cite{BannaiIto} or ~\cite{DiF-Zub-Trieste}).
The situation is similar to what happens for homogeneous spaces and 
reductive pairs of Lie algebras, but in the present case we are
in an associative (and commutative) algebra. A better analogy comes 
immediately to mind when we compare the representations
of $SO(3)$ and of $SU(2)$: All representations of the former are 
representations of the later, the set of irreps of $SO(3)$ is closed
under tensor products, and the coupling of an integer spin with a 
half (odd)-integer spin can be decomposed
on half-integer spins. Since $SO(3) = SU(2)/\ZZ_2$, we
 could say, by analogy, that the `quantum space' dual to the fusion algebra
of the graph $A_3$ is a quotient of the `quantum space' dual to the 
fusion algebra of the graph $E_6$.

\subsection{Paths on the $E_6$ graph}
An elementary path on a graph is a path, in the usual sense, starting 
at some vertex and ending at some other (or at the same)
vertex. Its length is counted by the number of edges entering in the 
definition of the path. Of course, paths can backtrack.
A (general) path is, by definition, a linear combination of 
elementary paths. The 
vector space of paths of length $n$
originating from $\sigma_i$, ending on $\sigma_j$ will be called 
$Path_{ij}^{n}$. 

The fusion graph of the group $SU(2)$ is an infinite half-line 
$A_\infty$, with vertices labelled by representations
$\sigma_0 = [1], \sigma_1 = [2], \ldots, \sigma_{2s} = [2s+1], 
\ldots$, where
the integers $1,2,3\ldots, d_s= 2s+1$ are 
 the dimensions of the irreducible representations of spin $s$. 
 Tensor multiplication by the
(two dimensional) fundamental representation is indeed such that 
$[d_s]\otimes [2] = [d_s-1] \oplus [d_s+1].$
An elementary path of length $n$ starting at the origin (the trivial 
representation) and ending on some $d_s$ 
can be put in one-to-one correspondance with a projector that 
projects the representation $[2]^n$ on the irreducible representation
$[d_s]$. The same comment can be made for discrete subgroups of 
$SU(2)$, for instance any binary polyhedral group, in which case 
the graph of fusion is given by the affine Dynkin diagrams 
$E_6^{(1)}$,  $E_7^{(1)}$,  $E_8^{(1)}$.
In the case of `genuine' Dynkin diagrams (not affine), we do not have
a group theoretical interpretation 
but the situation is similar.

Call $P_n = G. P_{n-1}$, where $G= N_1$ is, as usual, the adjacency 
matrix of the graph, 
and where $P_0= (1,0,0,0,0,0)^{T}$ (in the case of $E_6$). 
Clearly, the various components of the vector $P_n$ give the number 
of 
paths of length $n$,
starting at the origin ($\sigma_0$) and ending on the vertex 
corresponding to the chosen component.
 The following picture, 
 a kind of truncated Pascal triangle,  can therefore be generated 
very simply by
considering successive powers of the matrix $N_1$ acting on the 
(transpose) of
the vector $(1,0,0,0,0,0)$. 

Warning: in this picture, we have chosen the vertex order $0 1 2 3 5 
4$ rather than $0 1 2 5 4 3$ for aesthetical reasons;
for instance, one finds $P_7 = (0,21,0,20,0;15)$ but it is displayed 
as $(0,21,0,15,20,0)$.
We have therefore $21+15+20=56$ paths of length $7$ starting at the 
leftmost vertex on the graph $E_6$,
$21$ of them end on the vertex $\sigma_1$, $15$ end on $\sigma_3$ and 
$20$ end on $\sigma_5$.

Notice that the  picture stabilizes after a few steps:
 the whole structure appearing below
  can be graphically generated 
from the folded $E_6$ graph appearing at steps  $n=4$ and $5$ (the 
Bratteli diagram)
by  reflection and repetition 
down to infinity. This structure will be understood, at the end of 
this section, in terms of a 
tower of algebras (inclusions).


\vspace{1. cm}
\begin{figure}[h]
\unitlength 0.6mm
\begin{center}
\begin{picture}(120,125)

\put(5,120){\makebox(0,0){$ \ast $}}
\thicklines
\put(0,120){\line(1,-1){20}}

\put(25,100){\makebox(0,0){$ 1 $}}
\thinlines
\put(0,80){\line(1,1){20}}
\thicklines
\put(20,100){\line(1,-1){20}}

\put(5,80){\makebox(0,0){$ 1 $}}
\put(45,80){\makebox(0,0){$ 1 $}}
\thinlines
\put(0, 80){\line(1,-1){20}}
\thicklines
\put(40,80){\line(1,-2){10}}
\put(40,80){\line(1,-1){20}}
\thinlines
\put(20,60){\line(1,1){20}}

\put(25,60){\makebox(0,0){$ 2 $}}
\put(65,60){\makebox(0,0){$ 1 $ }}
\put(55,60){\makebox(0,0){$ 1 $}}

\put(0,40){\line(1,1){20}}
\put(40,40){\line(1,1){20}}
\put(40,40){\line(1,2){10}}

\put(20,60){\line(1,-1){20}}
\thicklines
\put(60,60){\line(1,-1){20}}
\thinlines
\put(5,40){\makebox(0,0){$ 2 $}}
\put(45,40){\makebox(0,0){$ 4 $}}
\put(85,40){\makebox(0,0){$ 1 $}}

\put(0,40){\line(1,-1){20}}
\put(40,40){\line(1,-2){10}}
\put(40,40){\line(1,-1){20}}

\put(20,20){\line(1,1){20}}
\put(60,20){\line(1,1){20}}

\put(25,20){\makebox(0,0){$ 6 $}}
\put(55,20){\makebox(0,0){$ 4 $}}
\put(65,20){\makebox(0,0){$ 5 $ }}

\put(0,0){\line(1,1){20}}
\put(40,0){\line(1,1){20}}
\put(40,0){\line(1,2){10}}

\put(20,20){\line(1,-1){20}}
\put(60,20){\line(1,-1){20}}

\put(0, - 5){\makebox(0,0){$ 6 $}}
\put(40, - 5){\makebox(0,0){$ 15 $}}
\put(80,- 5){\makebox(0,0){$ 5 $}}

\thicklines

\put(0, 0){\line(1,-1){20}}
\put(40, 0){\line(1,-2){10}}
\put(40, 0){\line(1,-1){20}}

\put(20, -20){\line(1,1){20}}
\put(60,-20){\line(1,1){20}}

\put(25,-20){\makebox(0,0){$ 21 $}}
\put(55,-20){\makebox(0,0){$ 15 $}}
\put(65,-20){\makebox(0,0){$ 20 $ }}

\thinlines

\put(0,-40){\line(1,1){20}}
\put(40,-40){\line(1,1){20}}
\put(40,-40){\line(1,2){10}}

\put(20,-20){\line(1,-1){20}}
\put(60,-20){\line(1,-1){20}}

\put(0, - 35){\makebox(0,0){$ 21 $}}
\put(40, - 35){\makebox(0,0){$ 56 $}}
\put(80,- 35){\makebox(0,0){$ 20 $}}

\put(0, 140){\makebox(0,0){$ \sigma_0 $}}
\put(25,140){\makebox(0,0){$ \sigma_1 $}}
\put(40, 140){\makebox(0,0){$ \sigma_2$}}
\put(55,140){\makebox(0,0){$\sigma_3 $}}
\put(65,140){\makebox(0,0){$ \sigma_5 $ }}
\put(75, 140){\makebox(0,0){$ \sigma_4 $}}
\put(85, 140){\makebox(0,0){$ $}}

\put(120,140){\makebox(0,0){n}}
\put(120,120){\makebox(0,0){}}
\put(120,100){\makebox(0,0){1}}
\put(120,80){\makebox(0,0){2}}
\put(120,60){\makebox(0,0){3}}
\put(120,40){\makebox(0,0){4}}
\put(120,20){\makebox(0,0){5}}
\put(120,0){\makebox(0,0){6}}
\put(120,-20){\makebox(0,0){7}}
\put(120,-40){\makebox(0,0){8}}
\end{picture}
\end{center}
\vspace{1.8 cm}
\label{fig: graphE6folded}
\end{figure}



In the group case $SU(2)$ for example, the vector $P_n$ has
infinitely many components (almost all are zero)
 and the value of the
component $P_n(d_s)$ gives the multiplicity of the 
representation $[d_s]$ in the $n$-th tensor power of the fundamental.
In that case, the semi-simple matrix algebra $TL(n)$ defined as a sum 
of simple blocks
 of dimensions  $P_n(d_s)$ is nothing else than the centralizer
algebra for the group $SU(2)$, also called ``Schur algebra'' or 
``Temperley-Lieb-Jones algebra'' (for the index $1/4$), and is a
well known quotient of the group algebra of the permutation group 
$S_n$;
the dimension of $TL(n)$, in that case, is given by the Catalan 
numbers $1,2,5,14,\ldots$.
When the graph is $E_6$, rather than $A_\infty$, there is no group 
theoretical interpretation but the construction of the commutant is 
similar
(it is a particular case of the Jones tower construction) 
and what we are describing  here is the path model for (the analogue) 
of a centralizer algebra. 
Still in the case of the graph $E_6$, we see on the previous picture 
that, for example, 
the algebra $TL(7)$ is isomorphic with $M(21,\CC)\oplus M(15,\CC) 
\oplus M(20,\CC)$ and is of dimension $21^2+15^2+20^2$.

\section{Essential paths and  essential matrices}
\subsection{Essential paths}
Let us start with the case of $SU(2)$. The consecutive $n$-th tensor 
powers of the fundamental representation $[2]$ can be decomposed
into irreducible representations ($[2]^2=[1]+[3]$, $[2]^3=2[2]+[4]$, 
$[2]^4=2[1]+ 3[3]+[5]$, \ldots). 
A given irreducible representation of dimension $d$ appears for the 
first time in the decomposition of $[2]^{d-1}$ and corresponds
to a particular projector in the vector space $(\CC^2)^{\otimes d-1}$ 
which is totally symmetric and therefore projects on the
space of symmetric tensors. These symmetric tensors provide a basis 
of this particular representation space and are, of course, in
one to one correspondance with symmetric polynomials in two complex 
variables $u, v$ (representations of given degree).
 From the point of view of paths, these
representations (projectors) correspond to non-backtracking paths of 
length $d-1$ starting at the origin (walking to the
right on the graph $A_\infty$). However, irreducible
representations of dimension $d$ appear not only in the reduction of 
$[2]^{d-1}$ but also in the reduction of $[2]^f$, when 
$f = d+1, d+3, \dots$. These representations are equivalent with the 
symmetric representations already described but they are
nevertheless distinct, as explicit given representations; 
the associated projectors are not symmetric and correspond to paths 
on $A_\infty$ that can backtrack.
The notion of ``essential path'', due to A. Ocneanu formalizes and 
generalizes the above remarks.
In the case of $SU(2)$, essential paths from the origin are just 
non-backtracking-right-moving paths starting from the origin 
($\tau_0 = [1]$) of $A_\infty$. There
is a one-to-one correspondance between such paths and irreducible 
symmetric representations. Clearly, ``essentiality''
is a meaningful property for a path or a projector, but a given {\sl explicit} 
irreducible  
representation, associated with an essential path, 
may very well be equivalent to another explicit representation 
which is still irreducible (of course, by definition of equivalence), 
but 
which is not associated with an essential path.
For instance the representation $[3]$ that appears in the reduction 
of $[2]^2$ 
corresponds to an essential path (starting from the origin), but the three equivalent 
representations $[3]$ 
that appear in the reduction of $[2]^4$ do not correspond to 
such paths. More generally, essential paths of length $n$ starting at 
a given irreducible representation of dimension $[a]$ (not necessarily 
the identity) correspond to projectors 
appearing in the decomposition of $[a]\otimes [n+1]$ into irreducible 
summands (here $[a]$ and $[n+1]$ are explicitly realized in terms of 
symmetric representations\footnote{The author acknowledges 
interesting comments by A. Garcia and R. Trinchero about this topic.}).

When we move from the case of $SU(2)$ to the case of finite subgroups 
of $SU(2)$, in particular the binary polyhedral
groups whose representation theory is described by the affine Dynkin 
diagrams $E_6^{(1)}$, $E_7^{(1)}$ and $E_8^{(1)}$,
the notion of essential paths can be obtained very simply by 
declaring that a path on the corresponding diagram
is essential if it describes an irreducible representation that 
appears in the branching 
of a symmetric representation of $SU(2)$ with respect to the chosen 
finite subgroup. A novel feature of these essential
paths is that they can backtrack (in general, such paths will be a 
linear combination of elementary paths). Essential
paths for the finite subgroups of $SU(2)$ can be of arbitrary length 
since symmetric representations of $SU(2)$ can
be of arbitrary degree (horizontal Young diagrams with an arbitrary 
number of boxes).

In more general situations (like the $E_6$ case which is the example that we are 
following in this paper), we need a  general definition
that encompasses all the previous concepts and provides a meaningful 
generalization. This definition was given by
A. Ocneanu (several seminars in 1995) and published in 
~\cite{Ocneanu:paths} and is as follows.

Take a graph $\Gamma$ described by an adjacency matrix $G$.
In this paper, edges of the graph are not oriented (as for the 
$ADE$); we may replace every unoriented edge
by a pair of edges with the same endpoints but carrying two opposite 
orientations. The notion of essential paths makes sense
for more general graphs -- for example those encoding the fusion by 
the two inequivalent fundamental representations of $SU(3)$
and its subgroups -- but we shall not be concerned with them in the 
present paper.

Call $\beta$ the norm of the graph $\Gamma$ (the biggest eigenvalue of
its  adjacency matrix) 
and  $D_{i}$ the components of the (normalized) Perron Frobenius eigenvector.  
Call $\sigma_{i}$ the vertices of $\Gamma$ and, if 
$\sigma_{j}$ is a neighbour of 
$\sigma_{i}$, call $\xi_{ij}$ the oriented edge
from $\sigma_{i}$ to $\sigma_{j}$. If $\Gamma$ is unoriented (the case for $ADE$
and affine $ADE$ diagrams), each edge should be considered  as carrying
both orientations.

An elementary path can be written either as a finite 
sequence of consecutive (\ie neighbours on the graph) vertices, 
$[\sigma_{a_1} \sigma_{a_2} \sigma_{a_3} \ldots ]$,
or, better, as a sequence $(\xi(1)\xi(2)\ldots)$ of consecutive edges, with
$\xi(1) = \xi_{a_{1}a_{2}}= \sigma_{a_1} \sigma_{a_2} $,
$\xi(2) = \xi_{a_{2}a_{3}} = \sigma_{a_2}  \sigma_{a_3} $, \etc.
Vertices are considered as paths of length $0$.

The length of the (possibly backtracking) path $( \xi(1)\xi(2)\ldots 
\xi(p) )$ is $p$.
We call $r(\xi_{ij})=\sigma_{j}$, the range of $\xi_{ij}$
and $s(\xi_{ij})=\sigma_{i}$, the source of $\xi_{ij}$.

For all edges $\xi(n+1) = \xi_{ij}$ that appear in an elementary path, 
we set  ${\xi(n+1)}^{-1} \doteq \xi_{ji}$.

For every integer $n >0$, the annihilation operator $C_{n}$,
acting on elementary paths of length $p$ is defined
as follows:  if $p \leq n$, $C_{n}$ vanishes, whereas if $ p \geq  n+1$ then
$$
C_{n} (\xi(1)\xi(2)\ldots\xi(n)\xi(n+1)\ldots) = 
\sqrt\frac{D_{r(\xi(n))}}{D_{s(\xi(n))}} 
\delta_{\xi(n),{\xi(n+1)}^{-1}}
 (\xi(1)\xi(2)\ldots{\hat\xi(n)}{\hat\xi(n+1)}\ldots) 
$$
Here, the symbol ``hat'' ( like  in $\hat \xi$) denotes omission.
The result is therefore either $0$ or a path of length $p-2$.
Intuitively, $C_{n}$ chops the round trip that possibly appears
at positions $n$ and $n+1$.

Acting on elementary path of length $p$, the creating operators $C^{\dag}_{n}$ are defined as follows:
if $n > p+1$, $C^{\dag}_{n}$ vanishes and, if $n \leq p+1$ then,
setting $j = r(\xi(n-1))$, 
$$
C^{\dag}_{n} (\xi(1)\ldots\xi(n-1)\ldots) = \sum_{d(j,k)=1}
\sqrt(\frac{D_{k}}{D_{j}})  (\xi(1)\ldots\xi(n-1)\xi_{jk}\xi_{kj}\ldots)
$$
The above sum is taken over the neighbours $\sigma_{k}$ of $\sigma_{j}$ on the graph.
Intuitively, this operator adds one  (or several) small round trip(s) 
at position $n$. 
The result is therefore either $0$ or a linear combination of paths of 
length $p+2$.

For instance, on paths of length zero (\ie vertices),
$$
C^{\dag}_{1} (\sigma_{j}) = \sum_{d(j,k)=1}
\sqrt(\frac{D_{k}}{D_{j}}) \xi_{jk}\xi_{kj} = \sum_{d(j,k)=1}
\sqrt(\frac{D_{k}}{D_{j}}) \, [\sigma_{j}\sigma_{k}\sigma_{j}]
$$

We already mentionned the fact that the Temperley-Lieb-Jones algebra 
$TL(n)$
could be constructed as  endomorphism algebra of the vector space of 
paths of length $n$ (path model).
The Jones' projectors $e_{k}$ are defined (as endomorphisms of 
$Path^n$) by 
$$
e_{k} \doteq \frac{1}{\beta} C^{\dag}_{k} C_{k} 
$$
The reader can indeed check that all Jones-Temperley-Lieb relations 
between the $e_i$
are verified. We remind the reader that
 $TL(n)$ is usually defined  as the $C^*$ algebra generated by 
$\{1,e_1,e_2,\ldots,e_{n-1}\}$
with relations 
\begin{eqnarray*}
e_{i}e_{i\pm 1}e_{i} &=& \tau e_{i} \cr
e_{i} e_{j}  &=& e_{j} e_{i} \mbox{\/ whenever \/} \vert i-j \vert 
\geq 2 \cr
e_{i}^{2}  &=& e_{i} 
\end{eqnarray*}
with $$\tau \doteq 1/\beta^{2}$$
We can now define what are ``essential paths'' for a general graph:
A path is called essential if it belongs to the intersection of the 
kernels 
of all the Jones projectors $e_{i}$'s (or if it belongs to 
the intersection of the kernels 
of all the anihilators $C_{i}$'s). The dedicated reader will show that 
this definition indeed
generalizes the naive definition given previously in the case of 
graphs
associated with $SU(2)$ and its subgroups.

The following difference of non essential paths of length $4$ starting at $\sigma_{0}$ and ending at $\sigma_{2}$
is an essential path of length $4$  on $E_{6}$:
$$\sqrt{[2]} (\xi_{01}\xi_{12}\xi_{23}\xi_{32})  - \sqrt \frac{[3]}{[2]} (\xi_{01}\xi_{12}\xi_{25}\xi_{52}) 
= \sqrt{[2]}  [0,1,2,3,2]- \sqrt \frac{[3]}{[2]}[0,1,2,5,2]$$
Here the brackets denote the $q$-numbers: $[2] = \frac{\sqrt 2}{\sqrt 3 -1}$ and
$[3] =
\frac{2}{\sqrt 3 -1}$. 
 
The Wenzl projector $p_{n}$ is, by 
definition, the projector that
take arbitrary paths of length $n$ and project them on the 
vector subspace of essential paths (call $EssPath^n$ this vector 
subspace). The original
definition of these projectors did not use the path model, but this 
equivalent definition
will be enough for our purpose.
In the case of $SU(2)$, those elements of Jones' 
algebra corresponding to 
projectors of $[2]^n$ on symmetric irreducible representations are Wenzl projectors, and
each Wenzl projector of that type is associated with an essential path from the 
origin. In the case of $SU(2)$, there
is only one symmetric representation in any dimension : the space of
essential paths of length $n$ starting at the origin of the graph 
(identity representation) is  one-dimensional and the map 
projecting the 
whole space of paths $Path^n $ on this one-dimensional space is the 
Wenzl projector $p_{n}$. It is of rank one. 
In the case of finite subgroups of $SU(2)$ or in the ``quantum'' 
cases corresponding to graphs $ADE$, 
this is not so : the space $EssPath^{n}$ is generally not of 
dimension 
$1$; for instance the $4$-dimensional irreducible representation of 
$SU(2)$
(which appears for the first time in $[2]^3$, and it can be associated with a 
single essential path of length $3$ starting at the
identity representation) can be decomposed into two irreps $[2']+[2'']$
of its binary tetrahedral subgroup; the Wenzl projector is of rank 
$2$.

\subsection{Dimension of $EssPath$. Essential matrices}

The only motivation for the previous discussion was to put what 
follows in its proper context.
Indeed, we shall not need, in this paper, to manipulate explicitly 
essential paths themselves (the
interested reader can do it, by using the previous general 
definitions). What we want to do here
 is only to give a simple method to count them; the method will be 
explicitly 
illustrated for the $E_6$ graph.

The main observation is that the dimension of the space of essential 
paths of length $n+1$ starting 
at $a$ and ending at $b$ is given by

$$ dim \, EssPath_{a,b}^{(n+1)} = dim (H_{n+1}) - dim \, 
EssPath_{a,b}^{(n-1)}$$
where $H_{n+1}$  is the space of linear combinations of paths of 
length $n+1$ which are essential on their first $n$ segments.
This result was obtained by A. Ocneanu (see also the lectures by J.B. 
Zuber \cite{Zuber:Bariloche}). The dimensions of spaces of essential 
paths can be encoded by a set of rectangular matrices that we shall 
call ``essential matrices'', but they are also called 
``intertwiners'' in other contexts like conformal field theory or 
statistical mechanics lattice models.

\smallskip

Once an (arbitrary) ordering of the vertices of the graph has been 
chosen -- so that we know how to associate a
positive integer (say $a$) to any chosen vertex --  
it 
is easy to show
 that the number  of essential paths of length $n$ starting at some 
vertex $a$
 and ending on the vertex $b$ is given by $b$-th component (for the 
 chosen vertex ordering) of the 
row vector $E_n(a)$ defined as follows:
\begin{itemize}
\item $E_a(0)$ is the  (line) vector caracterizing the chosen initial 
vertex,
\item $E_a(1) = E_a(0).G$ 
\item $E_a (n) = E_a(n-1).G - E_a (n-2)$
\end{itemize}
This is a kind of moderated Pascal rule: the number of essential 
paths 
(with fixed origin) of length $n$ reaching a particular vertex
is obtained from the  sum of number the paths of length $n-1$ 
reaching the neighbouring points (as in Pascal rule) 
by substracting  the number of paths of length $n-2$ 
reaching the chosen vertex.

In the case of $E_6$, we  order\footnote{Warning: again, the $E_{6}$-indices 
used to label matrices always refer to the ``name'' of the chosen vertices and not to the 
integer that labels corresponding rows or columns.}
the six vertices $\sigma_0, \sigma_1, 
\sigma_2, \sigma_5, \sigma_4, \sigma_3$ as before, so
that $E_0(0)=(1,0,0,0,0,0)$, $E_1(0)=(0,1,0,0,0,0)$, 
$E_2(0)=(0,0,1,0,0,0)$, $E_5(0) = (0,0,0,1,0,0)$,
$E_4(0)=(0,0,0,0,1,0)$, $E_3(0)=(0,0,0,0,0,1)$.

Starting from $E_a(0)$, one obtains in this way six rectangular 
matrices $E_a$ with infinitely many rows (labelled by $n$)
 and six columns (labelled by $b$).
The reader can check that all $E_a(n)$ are positive integers {\sl 
provided} $0 \leq n \leq 10$, but
this ceases to be true, as soon as $n>10$: 
 $E_0(11) = (0,0,0,0,0,0)$,  $E_0(12) = (0,0,0,0,-1,0)$ \dots
 
 We shall call ``essential matrices''  the six rectangular $11 \times 
6$ matrices
obtained by keeping only the first $11$ rows of the $E_a(.)'s$, and 
these finite dimensional rectangular matrices will still be denoted 
by $E_a$.
The particular matrix $E_{0}$, interpreted as an intertwiner, was obtained long ago (see for instance 
\cite{Pasquier},  \cite{Pearce-Zhou}); it is indeed easy to check that
it intertwines the adjacency matrices $G^{E_{6}}$ and $G^{A_{11}}$ of 
the Dynkin diagrams $E_{6}$ and $A_{11}$:
$$
E_{0}. G^{E_{6}} = G^{A_{11}} . E_{0}
$$

For all ADE graphs, the number of rows of essential matrices
is always given by the (dual) 
Coxeter number of the graph minus one.
In the case of $E_6$ this number is indeed $12-1=11$.
The components of the six rectangular matrix $E_a$ are denoted by 
$E_a[n,b]$. 

Once $E_{0}$ is known, one can obtain all others essential matrices thanks to the simple 
relation
$$
E_{a} = E_{0} N_{a}
$$

Rather than gathering all the information concerning essential paths 
on the diagram $E_6$ into a set
of six rectangular matrices $11\times 6$ (our essential matrices 
$E_a^{E_6}$'s), one can also define a set of eleven
{\sl square} matrices $6\times 6$ sometimes called ``fused adjacency matrices''
$$F_n^{E_6}[a,b] \doteq E_a^{E_6}[n,b]$$
In terms of the $F$ matrices, the definition given for the $E$ 
matrices reads simply $F_{0}  = 1, F_{1} = G$ and
$$
F_{n}F_{1} = F_{n-1} + F_{n+1}
$$

The important observation
is that these last matrices $F$ (relative to the Dynkin diagram 
$E_{6}$) build up a representation of the fusion 
algebra of the Dynkin diagram $A_{11}$;
this is clear since the previous relation is the usual 
 $SU(2)$ recurrence relation.
We shall come back to the 
relations between $E_{6}$ and $A_{11}$ in a later section.

\subsubsection{Essential matrices and essential paths for  $E_6$}

 The six essential matrices of $E_{6}$ are determined by straightforward 
 calculations using the previous recurrence relations. The reader can also
 easily obtain explicit expression for the corresponding eleven fused adjacency 
 matrices $F_{n}$. Here are the results for the rectangular matrices 
 $E_{a}$.

\smallskip

{\small

$$
\begin{array}{ccc}
E_0 = 
\left( \begin{array}{cccccc}
   1 & . & . & . & . & . \cr . & 
   1 & . & . & . & . \cr . & . & 
   1 & . & . & . \cr . & . & . & 
   1 & . & 1 \cr . & . & 1 & . & 
   1 & . \cr . & 1 & . & 1 & . & 
   . \cr 1 & . & 1 & . & . & . \cr
   . & 1 & . & . & . & 1 \cr . & 
   . & 1 & . & . & . \cr . & . & 
   . & 1 & . & . \cr . & . & . & 
   . & 1 & . \end{array} \right)

&

E_1 = 
\left( \begin{array}{cccccc}
   . & 1 & . & . & . & . \cr 1 & 
   . & 1 & . & . & . \cr . & 1 & 
   . & 1 & . & 1 \cr . & . & 2 & 
   . & 1 & . \cr . & 1 & . & 2 & 
   . & 1 \cr 1 & . & 2 & . & 1 & 
   . \cr . & 2 & . & 1 & . & 1 \cr
   1 & . & 2 & . & . & . \cr . & 
   1 & . & 1 & . & 1 \cr . & . & 
   1 & . & 1 & . \cr . & . & . & 
   1 & . & . \end{array} \right)

&

E_2 = 
\left( \begin{array}{cccccc}
   . & . & 1 & . & . & . \cr . & 
   1 & . & 1 & . & 1 \cr 1 & . & 
   2 & . & 1 & . \cr . & 2 & . & 
   2 & . & 1 \cr 1 & . & 3 & . & 
   1 & . \cr . & 2 & . & 2 & . & 
   2 \cr 1 & . & 3 & . & 1 & . \cr
   . & 2 & . & 2 & . & 1 \cr 1 & 
   . & 2 & . & 1 & . \cr . & 1 & 
   . & 1 & . & 1 \cr . & . & 1 & 
   . & . & . \end{array} \right)

\\
{} & {} & {} 
\\
E_5 = 
\left( \begin{array}{cccccc}
   . & . & . & 1 & . & . \cr . & 
   . & 1 & . & 1 & . \cr . & 1 & 
   . & 1 & . & 1 \cr 1 & . & 2 & 
   . & . & . \cr . & 2 & . & 1 & 
   . & 1 \cr 1 & . & 2 & . & 1 & 
   . \cr . & 1 & . & 2 & . & 1 \cr
   . & . & 2 & . & 1 & . \cr . & 
   1 & . & 1 & . & 1 \cr 1 & . & 
   1 & . & . & . \cr . & 1 & . & 
   . & . & . \end{array} \right)

&

E_4 = 
\left( \begin{array}{cccccc}
   . & . & . & . & 1 & . \cr . & 
   . & . & 1 & . & . \cr . & . & 
   1 & . & . & . \cr . & 1 & . & 
   . & . & 1 \cr 1 & . & 1 & . & 
   . & . \cr . & 1 & . & 1 & . & 
   . \cr . & . & 1 & . & 1 & . \cr
   . & . & . & 1 & . & 1 \cr . & 
   . & 1 & . & . & . \cr . & 1 & 
   . & . & . & . \cr 1 & . & . & 
   . & . & . \end{array} \right)

&

E_3 = 
\left( \begin{array}{cccccc}
   . & . & . & . & . & 1 \cr . & 
   . & 1 & . & . & . \cr . & 1 & 
   . & 1 & . & . \cr 1 & . & 1 & 
   . & 1 & . \cr . & 1 & . & 1 & 
   . & 1 \cr . & . & 2 & . & . & 
   . \cr . & 1 & . & 1 & . & 1 \cr
   1 & . & 1 & . & 1 & . \cr . & 
   1 & . & 1 & . & . \cr . & . & 
   1 & . & . & . \cr . & . & . & 
   . & . & 1 \end{array} \right)

\end{array}
$$
}

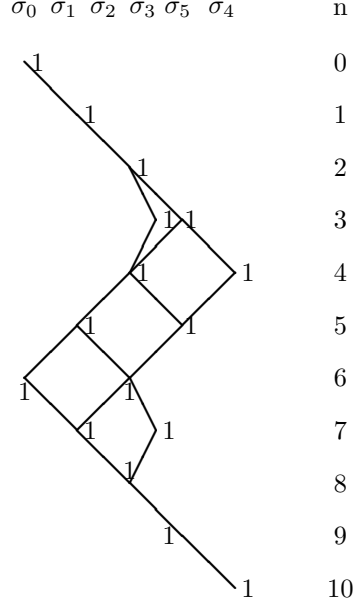
\begin{figure}[h]
\unitlength 0.35mm
\begin{center}
\begin{picture}(95,95)

\thicklines

\put(5,120){\makebox(0,0){$ 1 $}}

\put(0,120){\line(1,-1){20}}

\put(25,100){\makebox(0,0){$ 1 $}}


\put(20,100){\line(1,-1){20}}

\put(45,80){\makebox(0,0){$ 1 $}}


\put(40,80){\line(1,-2){10}}
\put(40,80){\line(1,-1){20}}


\put(65,60){\makebox(0,0){$ 1 $ }}
\put(55,60){\makebox(0,0){$ 1 $}}

\put(40,40){\line(1,1){20}}
\put(40,40){\line(1,2){10}}


\put(60,60){\line(1,-1){20}}

\put(45,40){\makebox(0,0){$ 1 $}}
\put(85,40){\makebox(0,0){$ 1 $}}

\put(40,40){\line(1,-1){20}}

\put(20,20){\line(1,1){20}}
\put(60,20){\line(1,1){20}}

\put(25,20){\makebox(0,0){$ 1 $}}
\put(65,20){\makebox(0,0){$ 1 $ }}

\put(0,0){\line(1,1){20}}
\put(40,0){\line(1,1){20}}

\put(20,20){\line(1,-1){20}}

\put(0, - 5){\makebox(0,0){$ 1 $}}
\put(40, - 5){\makebox(0,0){$ 1 $}}

\put(0, 0){\line(1,-1){20}}
\put(40, 0){\line(1,-2){10}}

\put(20, -20){\line(1,1){20}}

\put(25,-20){\makebox(0,0){$ 1 $}}
\put(55,-20){\makebox(0,0){$ 1 $}}

\put(40,-40){\line(1,2){10}}

\put(20,-20){\line(1,-1){20}}

\put(40, - 35){\makebox(0,0){$ 1 $}}

\put(40, -40){\line(1,-1){20}}
\put(60, -60){\line(1,-1){20}}

\put(55,-60){\makebox(0,0){$ 1 $}}
\put(85, -80){\makebox(0,0){$ 1 $}}

\put(0, 140){\makebox(0,0){$ \sigma_0 $}}
\put(15,140){\makebox(0,0){$ \sigma_1 $}}
\put(30, 140){\makebox(0,0){$ \sigma_2$}}
\put(45,140){\makebox(0,0){$\sigma_3 $}}
\put(60,140){\makebox(0,0){$ \sigma_5 $ }}
\put(75, 140){\makebox(0,0){$ \sigma_4 $}}
\put(90, 140){\makebox(0,0){$ $}}

\put(120,140){\makebox(0,0){n}}
\put(120,120){\makebox(0,0){0}}
\put(120,100){\makebox(0,0){1}}
\put(120,80){\makebox(0,0){2}}
\put(120,60){\makebox(0,0){3}}
\put(120,40){\makebox(0,0){4}}
\put(120,20){\makebox(0,0){5}}
\put(120,0){\makebox(0,0){6}}
\put(120,-20){\makebox(0,0){7}}
\put(120,-40){\makebox(0,0){8}}
\put(120,-60){\makebox(0,0){9}}
\put(120,-80){\makebox(0,0){10}}
\end{picture}
\vspace{3. cm}
\caption{Essential paths for $E_6$ starting from $\sigma_0$}
\label{fig: EssentialPathsGraphE0}
\end{center}
\end{figure}

It is clear that non zero entries of the  matrix $E_a$ encode also 
{\sl graphically} the structure of 
essential paths starting from $a$.For 
instance, we can ``read'', from the $E_0$ matrix, the following graph 
  (Fig. \ref{fig: EssentialPathsGraphE0}) giving all the essential 
paths leaving $\sigma_0$;
this  particular figure appears explicitly in
~\cite{Ocneanu:paths}. 
In this picture, for aesthetical reasons, the order of vertices was
chosen as $\sigma_0, \sigma_1, \sigma_2, \sigma_3, \sigma_5, 
\sigma_4$,
whereas the vertex order chosen for essential matrices was
$\sigma_0, \sigma_1, \sigma_2, \sigma_5, \sigma_4, \sigma_3$.


\vspace{1.0 cm}

\subsubsection{Essential paths and matrices for $A_{11}$}

The space of paths and space of essential paths can be defined for 
arbitrary ADE Dynkin diagrams (extended or not);
it is therefore natural to denote by  $E_a^X$ the essential
matrices relative to the choice of the graph $X$. Our previous 
results (called
$E_a$ in the case of the graph $E_6$) should therefore be denoted by 
$E_a^{E_6}$.

We shall not only need the essential matrices for the graph $E_6$ 
but also those for the graph $A_{11}$. 
The technique is exactly the same: first we call $\tau_0, 
\tau_1,\ldots, \tau_{10}$, from left to right, the vertices of 
$A_{11}$ 
and order them in a natural way;
then we build the adjacency matrix for $A_{11}$ that we may call 
again $G$ (but
this should be understood now as $G^{A_{11}}$ since we want to refer 
to this particular Dynkin diagram).
 We then build the associated graph fusion 
algebra and its matrix representation: we obtain in this way $11$ 
square matrices 
$N^{A_{11}}_n$ of size $11\times 11$. $N^{A_{11}}_0$ is the unit 
matrix,
$N^{A_{11}}_1= G^{A_{11}}$ is the generator, and the other fusion 
matrices are determined by the graph $A_{11}$: for 
$m+1 \leq 10$, we observe that
 $N^{A_{11}}_{m+1} = N^{A_{11}}_m.N^{A_{11}}_1 - N^{A_{11}}_{m-1};$ 
so that multiplication by $N^{A_{11}}_1$ therefore describes the usual coupling 
to spin $1/2$, but we have also $N^{A_{11}}_9 = N^{A_{11}}_1. N^{A_{11}}_{10}$. 
More generally, it is easy to prove (recurrence) the following formula 
(that can be interpreted in terms of couplings to higher spins):
$$
N^{A_{11}}_{n}.N^{A_{11}}_{m} = \sum_{p\in A_{11}} 
(N^{A_{11}}_{p})_{n,m} \, N^{A_{11}}_{p} 
$$
This last relation shows in particular that structure constants 
$C_{nmp}$ of the associative algebra $A_{11}$ can be identified with 
matrix elements $(N^{A_{11}}_{p})_{n,m}$.

One of the immediate consequences of the above is that, for $A_{11}$, and actually for 
all $A_N$ Dynkin diagrams,
there will be no difference between  graph fusion matrices,
fused adjacency matrices and
essential matrices, since
essential matrices in general (for any ADE graph) are precisely 
defined by the recurrence
formula that characterizes the fusion matrices of the $A_n$. In other 
words:
$$ N_m^{A_{11}} [i,j] =  E_m^{A_{11}} [i,j] =  F_m^{A_{11}} [i,j]$$
In the case of $A_{11}$, we obtain therefore eleven square $11\times 
11$ 
matrices of essential paths; this is to be contrasted with
 the case of $E_6$ where the six graph fusion matrices 
$N_a^{E_6} (b_1,b_2)$
are square ($6 \times 6$) but where the six
essential matrices $E_a^{E_6}(n,b)$ are rectangular ($11 \times 6$).

It is easy to show that the dimension of the space of essential paths 
of length
$n$, for $A_N$ Dynkin diagrams, $n \leq 10$,
 is $(N-n)(n+1)$. In the case of $A_{11}$, this dimension $d_n$ is

$$
\left( \begin{array}{cccccccccccc}
   n:      & 0 & 1 & 2 & 3 & 4 & 5 & 6 & 7 & 8 & 9 & 10 \cr
   d_n:& 11 & 20 & 27 & 32 & 35 & 36 & 35 & 32 & 27 & 20 & 11 
\end{array} \right)
$$
Notice that
$$
\sum_n d_n = 536 \qquad \sum_n (d_n)^2 = 8294
$$

From the above data, the reader may easily write down the
multiplication table for the fusion algebra and the eleven square 
matrices $(11\times 11)$ associated with
$A_{11}$; we shall not write them explicitly.

\subsection{Relations between $E_6$ and $A_{11}$}

The main relation defining the matrices $E_{a}$ for the $E_6$ diagram
can actually be understood
without having to use the notion of essential paths: written 
$E_{a}(n-1)G = E_{a}(n-2) + E_{a}(n)$, this relation describes the usual way of 
coupling irreducible representations of $A_{11}$ (a truncated version 
of $SU(2)$) and this observation leads to the (alreay defined) notion 
of ``fused adjacency matrices'', namely the eleven square matrices
$F_n^{E_6}[a,b] \doteq E_a^{E_6}[n,b]$. The known equality 
$E_{a}^{E_6}=E_{0}^{E_6}\, N_{a}^{E_6}$, when written in terms of $F$ 
matrices, leads to the
following relation between the six fusion graph matrices $N^{E_{6}}$ and 
the eleven fused adjacency matrices  $F^{E_{6}}$:
$$F^{E_{6}}_{n} = \sum_{c\in E_{6}} (F^{E_{6}}_{n})_{c0}\, N^{E_{6}}_{c}$$
Explicitly (we drop the $E_{6}$ label): $F_{0}=N_{0}, F_{1}=N_{1}, 
F_{2}=N_{2}, F_{3} = N_{3}+N_{5}, F_{4} = N_{2} + N_{4}, 
F_{5}= N_{1}+N_{5}, F_{6}=N_{0}+N_{2}, F_{7}=N_{1}+N_{3}, 
F_{8}=N_{2}, F_{9}=N_{5},
F_{10}=N_{4}$.

We prefer to use essential matrices
$E_{a}$'s but it is clear that the following discussion could also be carried 
out in terms of the  $F_n$ matrices. 

\subsubsection{$A_{11}$ labellings of the $E_6$ graph}

The six essential matrices $E_a$, or, equivalently, the eleven 
matrices $F_n$, or
 the six graphs of  essential paths, encode a
good deal of information. Let us consider for instance $E_0$; we see 
that, going from top to bottom,
and following essential paths starting from $\sigma_0$,
 after $0$-step, we are at $\sigma_0$, after $1$ step, we reach 
$\sigma_1$, after
$2$ steps, we reach $\sigma_2$, after $3$ steps, we reach either 
$\sigma_3$ or $\sigma_5$, \etc
Let us take another example, $E_2$; we see  that, starting from the 
vertex
$\sigma_2$, and following essential paths,  after $6$ steps (line 
$7=6+1$), we may reach
the vertex $\sigma_0$ (column $1$),  the vertex  $\sigma_2$ (column 
$3$) or the vertex $\sigma_4$;
notice that  
there are $3$ different 
ways to reach $\sigma_2$ (three linearly independent path of this 
type).

All these results just restate the fact that 
$E_{a,b}^{(n)} \doteq E_a[n,b]$ is the number of linearly
independent essential paths of length $n$ starting from $a$ and 
reaching $b$.

The index $n$ -- the length -- can be thought of as a label for a vertex of $A_{11}$,
but it can be also be understood as a (horizontal) Young diagram with 
$0 \leq n \leq 10$ boxes.

A particularly instructive way of illustrating these results is to 
give, 
for each chosen vertex chosen as initial point (marked with a star on 
the picture), a
graph of $E_6$ with the length of all possible essential paths 
indicated under the diagram.
In this way, we get immediately:

\begin{figure}[h]
\unitlength 0.6mm
\begin{center}
\begin{picture}(95,35)
\thinlines 
\multiput(25,10)(15,0){5}{\circle*{2}}
\put(55,25){\circle*{2}}
\put(25,12){$\ast$}
\thicklines
\put(25,10){\line(1,0){60}}
\put(55,10){\line(0,1){15}}

\put(25,3){\makebox(0,0){$0$}}
\put(25,-2){\makebox(0,0){$6$}}

\put(40,3){\makebox(0,0){$1$}}
\put(40,-2){\makebox(0,0){$5$}}
\put(40,-7){\makebox(0,0){$7$}}

\put(55,3){\makebox(0,0){$2$}}
\put(55,-2){\makebox(0,0){$4$}}
\put(55,-7){\makebox(0,0){$6$}}
\put(55,-12){\makebox(0,0){$8$}}

\put(70,3){\makebox(0,0){$3$}}
\put(70,-2){\makebox(0,0){$5$}}
\put(70,-7){\makebox(0,0){$9$}}

\put(85,3){\makebox(0,0){$4$}}
\put(85,-2){\makebox(0,0){$10$}}

\put(63,27){\makebox(0,0){$3,7$}}
\end{picture}
\bigskip
\caption{Essential Paths from 0}
\label{fig:EssPath0}
\end{center}
\end{figure}
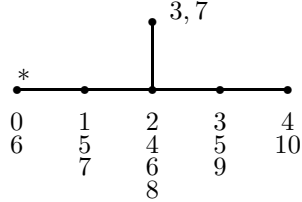
\begin{figure}[h]
\unitlength 0.6mm
\begin{center}
\begin{picture}(95,35)
\thinlines 
\multiput(25,10)(15,0){5}{\circle*{2}}
\put(55,25){\circle*{2}}
\put(40,12){$\ast$}
\thicklines
\put(25,10){\line(1,0){60}}
\put(55,10){\line(0,1){15}}

\put(25,3){\makebox(0,0){$1$}}
\put(25,-2){\makebox(0,0){$5$}}
\put(25,-7){\makebox(0,0){$7$}}

\put(40,3){\makebox(0,0){$0$}}
\put(40,-2){\makebox(0,0){$2$}}
\put(40,-7){\makebox(0,0){$4$}}
\put(40,-12){\makebox(0,0){$6_{2}$}}
\put(40,-17){\makebox(0,0){$8$}}

\put(55,3){\makebox(0,0){$1$}}
\put(55,-2){\makebox(0,0){$3_{2}$}}
\put(55,-7){\makebox(0,0){$5_{2}$}}
\put(55,-12){\makebox(0,0){$7_{2}$}}
\put(55,-17){\makebox(0,0){$9$}}

\put(70,3){\makebox(0,0){$2$}}
\put(70,-2){\makebox(0,0){$4_{2}$}}
\put(70,-7){\makebox(0,0){$6$}}
\put(70,-12){\makebox(0,0){$8$}}
\put(70,-17){\makebox(0,0){$10$}}

\put(85,3){\makebox(0,0){$3$}}
\put(85,-2){\makebox(0,0){$5$}}
\put(85,-7){\makebox(0,0){$9$}}

\put(68,27){\makebox(0,0){$2,4,6,8$}}
\end{picture}
\vspace{5 mm}
\bigskip
\caption{Essential Paths from 1}
\label{fig:EssPath1}
\end{center}
\end{figure}
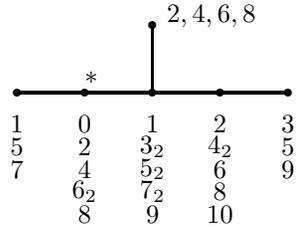

\centerline{\etc}
\bigskip

\begin{figure}
\includegraphics*{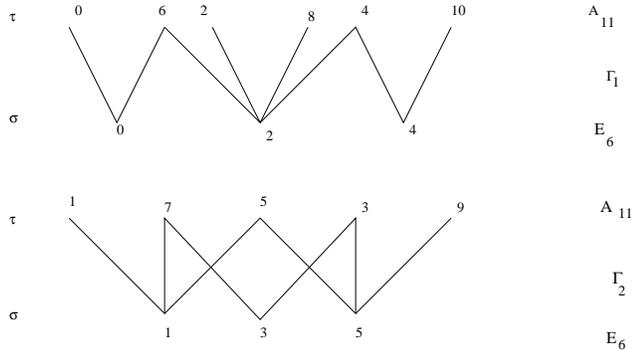}
\caption{Graphs relating $A_{11}$ and $E_6$}
\label{fig:principalgraphE6}
\end{figure}

We may look differently at the above correspondance(s) established 
between $A_{11}$ and $E_6$ by
displaying the results as figure \ref{fig:principalgraphE6}: 
Every  graph of the previous kind (choice of
the origin for the space of essential paths on $E_6$) gives rise to a 
new graph, where
the bottom line refers to $\sigma_a$ (the points of $E_6$), the top 
line 
to $\tau_j$ (the points of $A_{11}$) and there is a connecting line 
between the two, whenever
$\tau_j$ (actually the index $j$) appears below the vertex $\sigma_a$ 
in the figure describing essential paths from
a chosen vertex (figure \ref{fig:EssPath0}, \ref{fig:EssPath1}, 
\ldots).
For instance, if we choose the origin at $\sigma_0$ (essential paths 
from $0$),
we see that $6$ (denoting $\tau_6$) should be linked both to 
$\sigma_0$ (leftmost point) and $\sigma_2$ (middle point).
The graph gotten in this way may be disconnected; this is the case 
when we consider essential paths 
starting from $\sigma_0$ since we obtain a graph with two connected 
components
$\Gamma_1$ and $\Gamma_2$, namely figure  \ref{fig:principalgraphE6}.

Although we do not need this information here, it can be seen (cf. 
the last section)
 that the eleven points $ 0, \ldots 10$ of $A_{11}$ correspond to 
representations 
 of a finite dimensional quantum group (a quotient of $SU(2)_q$ when 
$q^{12} = \qch^{24} = 1$).

The two graphs $\Gamma_1$ and $\Gamma_2$ can also be used to describe 
the conformal embedding of the affine loop groups
$LSU(2)_{10} \subset LSO(5)_{1}$, but we shall not elaborate about 
this (\cite{Nahm}).
 Let us mention nevertheless that
 the basic $b$, vector $v$ and spinor $s$ representations of $LSO(5)$ 
obey the (Ising) fusion rules of the graph
$A_3$; we know  that $(b,s,v)$ have the same fusion rule as the 
$A_3$ fusion subalgebra of $E_6$ generated by $(\sigma_0, \sigma_3, 
\sigma_4)$.
For this reason we can also write $b \simeq \sigma_0 \rightarrow 
\tau_0 + \tau_6$, 
 $s \simeq \sigma_3 \rightarrow \tau_3 + \tau_7$ and  $v \simeq 
\sigma_4 \rightarrow \tau_4 + \tau_{10}$.
As it is well known, the same $A_3$ graph labels the three blocks of 
the modular invariant partition function relative to
the conformal field theory ($E_6$ case), a result
that we shall recover later when we deal with the toric matrices 
associated with the $E_6$ Dynkin diagram.

One may understand the above results
in terms of an analog of group theoretical elementary 
induction-restriction
of representations:
If, instead of the graphs $A_{11}$ and $E_6$, we were considering a 
finite
group $G$ together with a finite subgroup $H$, then, to each 
irreducible
representation $\sigma$ of $H$ on a vector space $V$
we could associate a vector bundle $G \times_{\sigma} V$ above the 
quotient
space $G/H$, and the spaces $\Gamma(G \times_{\sigma} V)$
of corresponding sections could be 
decomposed
in terms of irreducible representations of $G$ (induction). Therefore,
to each irreducible representation $\sigma$ of $H$, one can associate 
a (finite) list of numbers, namely the dimensions of the irreducible 
representations of $G$ appearing in this decomposition. 
This is precisely the classical analog of 
what is decribed by figure  \ref{fig:EssPath0}. Notice that when the 
chosen representation
$\sigma$ of $H$ is the trivial representation (the analog of the 
$\sigma_{0}$ vertex, the marked end point of our Dynkin diagram), the 
space of sections is nothing else than the space of functions over the homogeneous 
space $G/H$. The dedicated reader can easily work out this example 
when $G = \tilde I $ is the binary icosahedral group and $H= \tilde T$, 
is a binary tetrahedral 
subgroup; the corresponding induction - restriction theory is then described 
by the affine Dynkin diagram $E_{6}^{(1)}$ (replacing figure  
\ref{fig:EssPath0}) which is then ``decorated'' with the 
dimensions\footnote{This pedagogical calculation has been worked out in 
\cite{Coquereaux:ClassicalTetra}, after completion of the present article} of 
representations of $\tilde I$.

Another interpretation for the $A_{11}$ labelling of the $E_{6}$ graph  (and 
generalizations) can be obtained in the theory of induction-restriction
of sectors, applied to conformal field theory, see  ~\cite{Evans}.

\subsubsection{The $ E_6 \times E_6 \mapsto A_{11}$ table}

\label{sec: EEA}
The previous information can be also gathered in the following table 
 which can be directly read from
the essential matrices (for instance the fourth column of $E_1$
 (refering to vertex $\sigma_5$ of the graph $E_6$) has entries
$0,0,1,0,2,0,1,0,1,0,1$ refering to vertices $\tau_2, \tau_4, \tau_6, 
\tau_8, \tau_{10}$ of the graph $A_{11}$.
$$
\begin{array}{||c||cccccc}
E_a \times \widetilde E_b & 0 & 3 & 4 & 1 & 2 & 5 \\
\hline
\hline
0 & 0 6 & 3 7 & 4, 10 & 1 5 7 & 2 4 6 8 & 3 5 9 \\
3 & 3 7 & 0 4 6, 10 & 3 7 & 2 4 6 8 & 1 3 5 7_2 9 & 2 4 6 8 \\
4 & 4, 10 & 3 7 & 0 6 & 3 5 9 & 2 4 6 8 & 1 5 7 \\
1 & 1 5 7 & 2 4 6 8 & 3 5 9 & 0 2 4 6_2 8 & 1 3_2 5_2 7_2 9 & 2 4_2 6 
8,  10 \\
2 & 2 4 6 8 & 1 3 5_2 7 9 & 2 4 6 8 & 1 3_2 5_2 7_2 9 & 0 2_2 4_3 6_3 
8_2, 10 & 1 3_2 5_2 7_2 9 \\
5 & 3 5 9 & 2 4 6 8 & 1 5 7 & 2 4_2 6 8, 10 & 1 3_2 5_2 7_2 9 & 0 2 4 
6_2 8 
\label{TableAExEA=AA}
\end{array}
$$

Multiplying a matrix $E_a$ (of dimension $(11,6)$) by the transpose 
of a matrix $E_b$ (of dimension $(6,11)$)
gives a matrix $(11,11)$ and it can be checked that the above table 
gives the decomposition of a product of essential matrices of $E_{6}$
$E_a^{E_6}.\widetilde E_b^{E_6}$ in terms of  matrices of $A_{11}$ 
(remember that for $A_{11}$, and in 
general for $A_{N}$ graphs, we have equality between fused adjacency 
matrices, graph fusion matrices and essential matrices: $F_n^{A_{11}} =$ 
$E_n^{A_{11}} = N_n^{A_{11}}$).
For instance 
\begin{eqnarray*}
    E_1^{E_6}.\widetilde E_5^{E_6} & = & E_2^{A_{11}} + 2 
E_4^{A_{11}} + E_6^{A_{11}} + E_8^{A_{11}} + E_{10}^{A_{11}} \\
{} & = & N_2^{A_{11}} + 2 
N_4^{A_{11}} + N_6^{A_{11}} + N_8^{A_{11}} + N_{10}^{A_{11}}
\end{eqnarray*}

 The general proof uses the fact that 
the (symmetric) $F^{E_{6}}$ matrices constitute a representation of the $A_{11}$ 
 algebra and that structure constants of this algebra are given by 
 matrix elements of the fusion graph matrices themselves; it goes as 
 follows:
$(E_a^{E_6}.\widetilde E_b^{E_6})_{mn} = \sum_{c\in E_{6}}(E_a^{E_6})_{mc}.(\widetilde 
E_b^{E_6})_{cn}=
\sum_{c\in E_{6}}(F_m^{E_6})_{ac}.(F_n^{E_6})_{cb} = \sum_{p\in A_{11}}
C_{mnp} (F_p^{E_6})_{ab} =  \sum_{p\in A_{11}} (E_a^{E_6})_{pb} 
(N_{p}^{A_{11}})_{m,n}.$

Therefore 
$E_a^{E_6}.\widetilde E_b^{E_6} = \sum_{p\in A_{11}} (E_a^{E_6})_{pb} 
N_{p}^{A_{11}}.$
The same relation could be also written in a more symmetric way as
$$E_a^{E_6}.\widetilde E_b^{E_6} = \sum_{p\in A_{11}} (F_p^{E_6})_{ab} \,
F_{p}^{A_{11}}$$

\subsubsection{Invariants and para-invariants}
Following the terminology due to A.Ocneanu, a para-invariant of 
degree $n$, 
relative to the vertex $x$, and denoted $E_{x,x}^{(n)}$, 
 is an essential path of lenth $n$
starting at $x$ and coming back at $x$.
 When $x = 0$ 
we get an invariant in the usual sense. Notice that the index $n$ can 
be thought of a particular kind of Young diagram made of a single
horizontal row  with $n$ boxes.
$$
\left( \begin{array}{cccccccccccc}
   n:         & 0 & 1 & 2 & 3 & 4 & 5 & 6 & 7 & 8 & 9 & 10 \cr
      {} & {} & {} & {} & {} & {} & {} & {} & {} & {} & {} \cr
E_{0,0}^{(n)}:& 1 & 0 & 0 & 0 & 0 & 0 & 1 & 0 & 0 & 0 &  0 \cr
      {} & {} & {} & {} & {} & {} & {} & {} & {} & {} & {} \cr
E_{1,1}^{(n)}:& 1 & 0 & 1 & 0 & 1 & 0 & 2 & 0 & 1 & 0 &  0 \cr
E_{2,2}^{(n)}:& 1 & 0 & 2 & 0 & 3 & 0 & 3 & 0 & 2 & 0 &  1 \cr
E_{5,5}^{(n)}:& 1 & 0 & 1 & 0 & 1 & 0 & 2 & 0 & 1 & 0 &  0 \cr
E_{4,4}^{(n)}:& 1 & 0 & 0 & 0 & 0 & 0 & 1 & 0 & 0 & 0 &  0 \cr
E_{3,3}^{(n)}:& 1 & 0 & 0 & 0 & 1 & 0 & 1 & 0 & 0 & 0 &  1 \cr
      {} & {} & {} & {} & {} & {} & {} & {} & {} & {} & {} \cr
      I^{(n)}:& 6 & 0 & 4 & 0 & 6 & 0 & 10 & 0 & 4 & 0 &  2 \cr
\end{array} \right)
$$

Here we call $I_n$ the total number of para-invariants of degree $n$.

An invariant of degree $n$ is an essential path of length $n$; 
starting 
at $\sigma_0$ (origin) and coming back to $\sigma_0$ (extremity). 
Their number is
$E_{0,0}^{(n)}$.
$$
\left( \begin{array}{cccccccccccc}
   n:         & 0 & 1 & 2 & 3 & 4 & 5 & 6 & 7 & 8 & 9 & 10 \cr
E_{0,0}^{(n)}:& 1 & 0 & 0 & 0 & 0 & 0 & 1 & 0 & 0 & 0 &  0 
\end{array} \right)
$$
These invariants (in the case of $E_6$ we have only one, in degree 
$6$)
 are quantum analog of the famous Klein invariants for polyhedra. 
Let us explain this:
Take a classical polyhedron, put its vertices on the sphere, make a 
stereographical projection,
build a polynomial that vanishes precisely at the location of the 
projected vertices (or center of faces, or mid-edges):
you get a polynomial which, by construction, is invariant under the 
symmetry group of the polyhedron (at least
projectively) since group elements only permute the roots.
 This is the historical method -- see in particular the famous little 
book ~\cite{Klein}.
In the case of the tetrahedron, for instance, you get the three 
polynomials (in homogeneous coordinates):
$V = u^4 + 2i\sqrt 3 u^2 v^2 + v^4$, $E = uv(u^4-v^4)$ and 
$F=u^4-2i\sqrt 3 u^2 v^2 + v^4$. Actually $V$ and $F$ are only
projectively invariant, but $X=108^{1/4} E $, $Y = - VF = -(u^8+v^8 + 
14 u^{4} v^{4})$ and $Z = V^3 - i X^2
=(u^{12}+v^{12})-33(u^8 v^4 + u^4 v^8)$ are (absolute) invariants, of 
degrees $6,8,12$.
 Together with the relation $X^4+Y^3+Z^2=0$, they generate the whole 
set of invariants.
Alternatively you can build the $p$-th power of the fundamental 
representation  (it is $2$-dimensional) of the symmetry group
of the chosen binary polyhedral group, and choose $p$ such that there 
exists one  essential path of length $p$ starting at the origin
of the graph of tensorisation by the fundamental representation 
(therefore one of the affine $ADE$ graphs) that returns to the origin.
Therefore you get a symmetric tensor (since the path is essential), 
hence a homogeneous polynomial of degree $p$; moreover this
polynomial is invariant since the path goes back to the origin (the 
identity representation). By calculating explicitly
the projectors corresponding to the (unique) essential path of 
$[2]^6$, $[2]^8$ and $[2]^{12}$ on the affine $E_6^{(1)}$ graph,
one can recover the
polynomials $X,Y,Z$. The reader may refer to the set of notes 
\cite{Coquereaux:ClassicalTetra} where this (tedious) calculation can
be found.

Returning to the quantum tetrahedron case (the $E_6$ Dynkin diagram), 
we do not have a polyhedron to start with\ldots Nevertheless,
we still have essential paths, so the above notion of invariants 
(defined as essential paths starting at the origin and returning
at the origin) makes sense. Notice that it would be nice to be able 
to exhibit a polynomial with non commuting
variables $u$ and $v$ manifesting some invariance with respect to an 
appropriate quantum group action. This was not obtained, so far.

\subsubsection{Diagonalisation of the fusion algebra of $A_{11}$}
\label{sec:A11}

The eleven  $11 \times 11$  square matrices $ N_n^{A_{11}}$ that we 
just introduced commute with one 
another, therefore, they can be simultaneously diagonalized: one can 
find a matrix $S$ which is such
that all the $S^{-1}. N_n^{A_{11}} . S$ are diagonal. This matrix, 
which is itself $11\times 11$, can
be considered as the  ``non commutative character table'' of $A_{11}$ 
(see the remark made at the end of section \ref{sec:E6}); it will be 
explicitly given in
section \ref{sec:Verlinde}.

\subsection{The algebra ${\cal A}$ of endormorphisms of essential 
paths}
\label{xiab}

The dimension of the vector space of essential paths of length $n$ 
(with arbitrary origin and extremity) is $d_n \doteq \sum_{a,b} 
E_{a,b}^{(n)}$ :
we take the sum of all matrix elements of the row $n+1$ of each 
matrix $E_a$ (since length $0$ corresponds to the
first row of the essential matrices), then we sum over $a$.

$$
\left( \begin{array}{cccccccccccc}
   n:      & 0 & 1 & 2 & 3 & 4 & 5 & 6 & 7 & 8 & 9 & 10 \cr
   d_n:& 6 & 10 & 14 & 18 & 20 & 20 & 20 & 18 & 14 & 10 & 6 
\end{array} \right)
$$

One may interpret $n$ as a length, or as a particular vertex 
($\tau_n$) of the $A_{11}$ graph.
An essential path $\xi$ of length $n$ from $a$ to $b$ can be denoted 
by $\xi_{a,b}^n$ and pictured as follows \label{fig:EEA}:
\includegraphics*{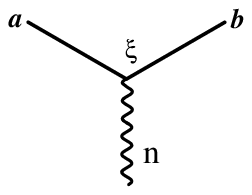}
We introduce one such vertex  $\xi_{a,b}^n$ whenever $n$ appears, in 
the
previous $E_a \times \widetilde E_b$ table, at the intersection of rows 
and columns $a$ and $b$ (in case
of multiplicity, one has to introduce different labels  $\xi_{a,b}^n, 
\zeta_{a,b}^n$, \ldots)
Therefore, we have  $6$ vertices of type $\xi_{\cdot \cdot}^0$,  
$10$ vertices of type $\xi_{\cdot \cdot}^1$,
$14$  of type $\xi_{\cdot  \cdot}^2$, \etc
and the dimension of $EssPath$ is
$$
\sum_{n} d_{(n)} = 156
$$

The vector space of essential paths is graded by the length; $EssPath 
= \bigoplus_n EssPath^n$.
Let us consider  ${\cal A}^n \doteq End \, EssPath^n  \simeq 
EssPath^n \otimes EssPath^n$, 
 the algebra of endomorphisms of this particular vector subspace.
Notice that  ${\cal A}^n$
 is isomorphic with an algebra of square
matrices of dimension $d_n^2$ where $d_n = dim \, EssPath^n$.
Let us also consider the graded algebra ${\cal A} \doteq  
\bigoplus_n  End \, EssPath^n$ and call it ``the algebra
 of endomorphisms of essential paths''.
${\cal A}$ is a direct sum of matrix algebras. Its dimension is
$\sum_n d_n^2$:
$$
dim({\cal A}) = 6^2 + 10^2 + 14^2 +18^2 + 20^2 + 20^2 + 20^2 + 18^2 + 
14^2 + 10^2 + 6^2 = 2512
$$

A first basis for the subalgebra  ${\cal A}^n$ is given by the tensor 
products $\xi\otimes \eta$ where $\xi$ and $\eta$ 
run in a basis of $EssPath^n$. Such tensor products can be described 
by the following picture
that looks like a 
diffusion graph in particle physics (dually, this is a double 
triangle).
\includegraphics{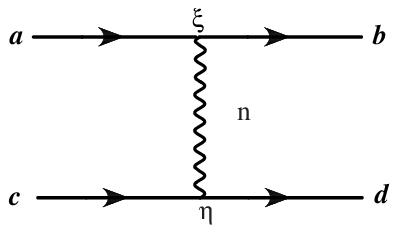}
 The dimension of ${\cal A}$ can be recovered by a simple 
exercise in combinatorics: the counting of
all possible labelled diffusion graphs.

\section{The algebra $E_6 \otimes_{A_3} E_6$}
\subsection{Definition}
In the present section, we introduce and discuss the properties
of the algebra $$S \doteq E_6 \otimes_{A_3} E_6$$ It will be called
``algebra of quantum symmetries''. Let us first explain this 
definition.
We start from the fusion algebra of $E_6$; this is a commutative 
algebra, and, in particular, a vector space. We may consider its 
tensor square $E_6 \otimes E_6$, which is a vector space of dimension 
$6^2$ and can be endowed with a
natural multiplication : $(a_1 \otimes b_1)(a_2 \otimes b_2) \doteq 
(a_1 a_2
\otimes b_1 b_2)$. Here the tensor product $\otimes$ is the usual 
tensor
product. In order to construct
the algebra $S$, we do not take the tensor 
product as  above, over
the complex numbers, but over the subalgebra $A_3$; this means that, 
given
$a$ and $b$ in the fusion algebra  $E_6$,
we identify $a x \otimes b \equiv a \otimes x b$ not only when $x$ is 
a complex 
number, but also when $x$ is an element of the subalgebra $A_3$ 
generated
by $\sigma_0, \sigma_4, \sigma_3$.
This tensor product will be denoted by $\stackrel{\cdot}{\otimes}$ 
rather than $\otimes$.
 We remember that $A_3$  is a very 
particular subalgebra of the (commutative) algebra $E_6$, on which it 
acts non trivially by multiplication. The dimension of the algebra 
$S$, just constructed, is therefore
not equal to $6^2$ but to $6^2/3 = 12$.

\subsection{A linear basis for $S$}

The following elements build up a set ($L \cup R \cup A \cup C$)
 of $12$ linearly independent generators for $S$.
Here and below, we write $a$ rather than $\sigma_a$.

$$ L = \{1\stackrel{\cdot}{\otimes} 0, 2 \stackrel{\cdot}{\otimes} 0, 
5 \stackrel{\cdot}{\otimes} 0 \}$$
$$ R = \{0 \stackrel{\cdot}{\otimes} 1, 0 \stackrel{\cdot}{\otimes} 2 
= 3 \stackrel{\cdot}{\otimes} 1, 0 \stackrel{\cdot}{\otimes} 5 = 4 
\stackrel{\cdot}{\otimes} 1\} $$
$$ A = \{0  \stackrel{\cdot}{\otimes} 0, 3 \stackrel{\cdot}{\otimes} 
0 = 0 \stackrel{\cdot}{\otimes} 3, 4 \stackrel{\cdot}{\otimes} 0 = 0 
\stackrel{\cdot}{\otimes} 4 \}$$
$$ C = \{ 1 \stackrel{\cdot}{\otimes} 1, 2 \stackrel{\cdot}{\otimes} 
1 = 1 \stackrel{\cdot}{\otimes} 2 , 5 \stackrel{\cdot}{\otimes} 1 
 = 1 \stackrel{\cdot}{\otimes} 5 
\}$$

For reasons that will be explained later, we have split this basis, 
made of twelve elements, into four
subsets: $L,R,A,C$.

Showing that these $12$ elements are linearly independent is 
straigtforward, and since $S$ is $12$-dimensional,
 we have a basis. What are not totally obvious are
 the above mentionned equalities; let us prove them.
The calculations use (of course) the multiplication table for the 
fusion algebra of the $E_6$ graph, 
and the fact that $0,3,4$ can ``jump'' over the tensor product sign 
$\stackrel{\cdot}{\otimes}$.

\begin{eqnarray*}
3 \stackrel{\cdot}{\otimes} 1 &=& 0 \stackrel{\cdot}{\otimes} 3.1 = 0 
\stackrel{\cdot}{\otimes} 2 \\
4 \stackrel{\cdot}{\otimes} 1 &=& 0 \stackrel{\cdot}{\otimes} 4.1 = 0 
\stackrel{\cdot}{\otimes} 5 \\
3 \stackrel{\cdot}{\otimes} 0 &=& 0 \stackrel{\cdot}{\otimes} 3.0 = 0 
\stackrel{\cdot}{\otimes} 3 \\
4 \stackrel{\cdot}{\otimes} 0 &=& 0 \stackrel{\cdot}{\otimes} 4.0 = 0 
\stackrel{\cdot}{\otimes} 4 \\
2 \stackrel{\cdot}{\otimes} 1 &=& 1.3 \stackrel{\cdot}{\otimes} 1 = 1 
\stackrel{\cdot}{\otimes} 3.1 = 1 \stackrel{\cdot}{\otimes} 2 \\
5 \stackrel{\cdot}{\otimes} 1 &=& 1.4 \stackrel{\cdot}{\otimes} 1 = 1 
\stackrel{\cdot}{\otimes} 4.1 = 1 \stackrel{\cdot}{\otimes} 5 
\end{eqnarray*} 
 The reader can prove, in the same way, many other identities, like 
for instance
$$2 \stackrel{\cdot}{\otimes} 2 = 5.3 \stackrel{\cdot}{\otimes} 2 = 5 
\stackrel{\cdot}{\otimes} 3.2 = 5 \stackrel{\cdot}{\otimes} (1+5) = 5 
\stackrel{\cdot}{\otimes} 1 + 5 \stackrel{\cdot}{\otimes} 5$$
$$ 5 \stackrel{\cdot}{\otimes}  5 = 1.4 \stackrel{\cdot}{\otimes} 5 = 
1 \stackrel{\cdot}{\otimes} 4.5 = 1 \stackrel{\cdot}{\otimes} 1 $$

\subsection{Structure and multiplication table of $S$ }

Using the previous technique,  one can build a multiplication table 
$12 \times 12$.
However, it is enough to observe the following:

The subalgebra $E_6 \stackrel{\cdot}{\otimes} 0$ of $S$ linearly generated by 
$L \cup A$ ie by the six elements
 $ a \stackrel{\cdot}{\otimes} 0$ 
(where $a$ runs in the set $\{0,3,4,1,2,5\}$)
is obviously isomorphic with the fusion algebra of the graph $E_6$ 
itself; it is called the ``chiral left 
subalgebra''.
We can make a similar remark for the ``chiral right subalgebra'' $0 
\stackrel{\cdot}{\otimes} E_6$ linearly generated by
$R \cup A$.
Since $1$ is the (algebraic) generator of the fusion algebra, we see 
that $1\stackrel{\cdot}{\otimes} 0$ and $0 \stackrel{\cdot}{\otimes} 
1$ separately generate (algebraically) the left and right subalgebras 
$E_6 \stackrel{\cdot}{\otimes} 0$ and $0 \stackrel{\cdot}{\otimes} 
E_6$.
The ``ambichiral subalgebra''is the intersection of left and right 
chiral subalgebras; it is  linearly generated by the elements of the 
set $A$ and is isomorphic with the algebra generated by $0,3,4$, \ie with 
the graph fusion algebra of  $A_3$.
Notice that the vector space linearly generated by the 
elements of $C$ (standing for ``Complement'') is not a subalgebra; 
a final observation is that the three basis vectors of $C$ can be 
obtained by multiplying elements of $L$ and
elements of $R$. Indeed:
\begin{eqnarray*}
1 \stackrel{\cdot}{\otimes} 1 &=& (0\stackrel{\cdot}{\otimes} 1)(1 
\stackrel{\cdot}{\otimes} 0) \\
2 \stackrel{\cdot}{\otimes} 1 &=& (3.1\stackrel{\cdot}{\otimes} 1) =  
(3\stackrel{\cdot}{\otimes} 1)(1 \stackrel{\cdot}{\otimes} 0) \\
5 \stackrel{\cdot}{\otimes} 1 &=& (4.1 \stackrel{\cdot}{\otimes} 1) = 
(4\stackrel{\cdot}{\otimes} 1)(1 \stackrel{\cdot}{\otimes} 0) 
\end{eqnarray*}

\smallskip

The conclusion is that $1 \otimesdot 0$ and $0 \otimesdot 1$ generate 
algebraically the algebra $S$.
These two elements are called left and right generators;
it is therefore enough to know the multiplication of arbitrary 
elements by these two generators
to reconstruct the whole multiplication table of $S$.


\subsection{Ocneanu graph of quantum symmetries}

The multiplication of arbitrary elements by the two (left and right) 
generators
can be best summarized by the corresponding Cayley graph.

\label{OcneanuGraph}
\includegraphics{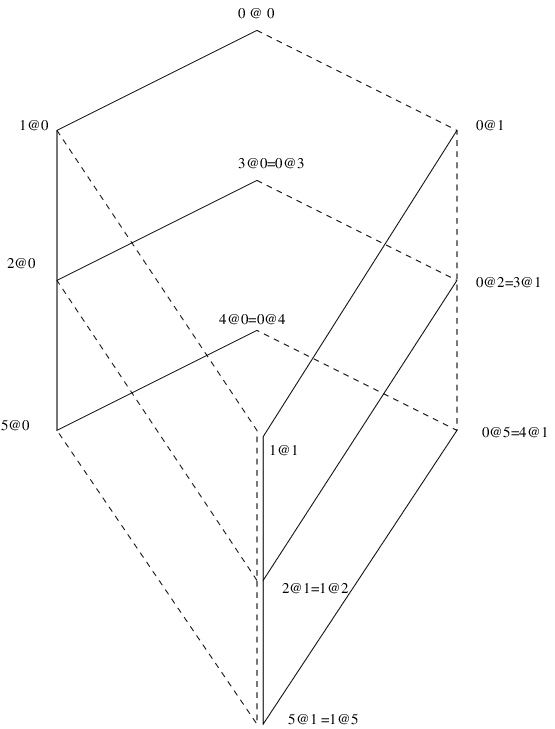}

Multiplication by $1\otimesdot 0$ is given by continuous lines and 
multiplication by $0\otimesdot 1$
is given by dotted lines. For instance, we read from this graph the 
equalities
\begin{eqnarray*}
(2 \otimesdot 1)(0 \otimesdot 1) & = &  1 \otimesdot 1 + 2 \otimesdot 
0 + 5 \otimesdot 1 \\
(5 \otimesdot 1)(1 \otimesdot 0) & = &  2 \otimesdot 1 + 0 \otimesdot 
5
\end{eqnarray*}

Let us prove for instance the first equality.
The left hand side is also equal to $(2.0 \otimesdot 1.1) = 2 
\otimesdot (0+2) = 2 \otimesdot 0 + 2 \otimesdot 2$,
but $2\otimesdot 2 = 5\otimesdot 1 + 1\otimesdot 1$, as shown 
previously; hence the result.

This graph was obtained by A. Ocneanu as a graph encoding the
quantum symmetries of $E_6$, defined in a totally different way 
(irreducible connections on a graph -- we 
shall come back to this
original definition in one of the appendices), and interpreted as a 
Cayley graph describing multiplication by two particular generators.
One of our observations, in the present paper,
is to notice that this algebra of 
quantum symmetries is isomorphic with the associative algebra  $E_6 \otimes_{A_3} E_6$.

In the case of $A_N$, the Ocneanu graph is obtained by setting $S = 
A_N \otimes_{A_N} A_N$ (there are $N$ points).
In the case of $E_8$, it is obtained by setting $S = E_8 
\otimes_{A_2} E_8$ (there are $8\times 8/2=32$ points).
The cases of $E_7$ and $D_{odd}$ are special since the fusion table 
of those Dynkin diagram 
cannot be constructed (they do
not define an hypergroup with {\sl positive} structure constants); 
this does not mean, of course that one cannot consider their quantum 
symmetries, but the technique
that we are explaining here should be adapted. The case of $D_{even}$ 
is also special because the two vertices that constitute
the ``fork'' of the graph do not behave like the others and give rise to an 
algebra of quantum symmetries which is not
commutative, contrarily to the other examples encountered so far (the 
number of points of the Ocneanu graph, for $D_{2n}$ is
$(2n-2)\times (2n-2) /(n-1) + 2^2 = 4n$ (the algebra itself being
isomorphic with $\CC^{4n-4} \oplus M(2,\CC)$).
The results themselves can be found in the paper \cite{Ocneanu:paths}, 
the details and proofs should appear in the work of A. Ocneanu, when available, or, 
following the techniques adapted from what is explained here,
in \cite{Coque-Gil:ADE} and in  part of the thesis \cite{Schieber:thesis}

\subsection{The $E_6\times E_6 \mapsto S$ table}
\label{sec: EES}
To each element of $S = E_6 \otimes_{A_3} E_6$ (for instance 
$\sigma_5 \otimesdot \sigma_1$), we may associate
one representative in $E_6 \otimes E_6$ (for instance $\sigma_5 
\otimes \sigma_1$). Then, we may apply the multiplication
map $a \otimes b \rightarrow ab$ to get one element in $E_6$ (here it 
is $\sigma_5 \sigma_1 = \sigma_2 + \sigma_4$).
The result is obviously independent on the choice of the 
representative since both $a x \otimes b$ and $a \otimes x b$, with
$x \in A_3$, have the same image $axb$ in $E_6$. Now we can represent 
the obtained element of $E_6$ ($\sigma_5 \sigma_1$ in our example) by 
the corresponding graph fusion matrix (namely $N_5 N_1$ in our example). 
The result is a $6\times 6$ matrix with rows
and columns labelled by the Dynkin
diagram of $E_6$. In our example,  using the (ordered) basis $012543$,
 we have the correspondance
$$
\sigma_5 \otimesdot \sigma_1 \rightarrow S_{51} \doteq
\left(\begin{array}{cccccc}
. & . & 1 & . & 1 & . \\
. & 1 & . & 2 & . & 1 \\
1 & . & 3 & . & 1 & . \\
. & 2 & . & 1 & . & 1 \\
1 & . & 1 & . & . & . \\
. & 1 & . & 1 & . & 1 
\end{array}\right)
$$
By this construction we obtain twelve $(6,6)$ matrices $S_{ab}$, each 
one being associated  with a particular basis element 
$a \otimesdot b$ of $S$.
We shall define $d_{a\otimesdot b} = \sum_{c,d}(S_{ab})_{cd}$, for 
instance
$d_{5\otimesdot 1}=20$.

The following table summarises the results: the element $a\otimesdot 
b$ of $S$ (denoted $ab$ in the table, to save space) appears at
intersection of $\widetilde E_c \times E_d$ (with possible multiplicity 
$m$) whenever the entry $(c,d)$ (labelling 
the $E_6$ graph) of the matrix $S_{ab}$
is equal to $m$. For instance the entry $51$ appears in the table
with multiplicity $3$ at the intersection of row $2$ and column $2$ 
since the matrix element of $S_{51}$ relative to the row and column indexed by $(\sigma_2, \sigma_2)$ is equal to $3$.

One may use these results  to define  a kind of generalized 
fusion law $\odot: E_6 \times E_6 \mapsto S$ (hence
the name of the section). In this way, one obtains the following 
particularly useful equality:
$$\sigma_0 \odot \sigma_0 = 0\otimesdot 0 + 1 \otimesdot 1$$

Here comes the table: 

{\tiny
$$
\begin{array}{||c||cccccc}
\widetilde E_c \times  E_d & 0 & 3 & 4 & 1 & 2 & 5 \\
\hline
\hline
0 & 00,11 & 30,21 & 40,51 & 10,01,21 & 20,11,31,51 & 50,21,41 \\
3 & 30,21 & 00, 40, 11, 51 & 30, 21 & 20, 11, 31, 51 & 10, 50, 01, 
21_2, 41 & 20, 11, 31, 51 \\
4 & 40,51 & 30,21 & 00,11 & 50,21,41 & 20,11,31,51 & 10,01,  21 \\
1 & 10,01,21 & 20,11,31,51 & 50,21,41 & 00, 20, 11_2, 31, 51 & 
10,30,50,01,21_3,41 & 20,40,11,31,51_2 \\
2 & 20,11,31,51 & 10,50,01,21_2,41 & 20,11,31,51 & 
10,30,50,01,21_3,41 & 00,20_2,40,11_3,31_2,51_3 &10,30,50,01,21_3,41\\
5 &50,21,41 & 20,11,31,51 & 10,01,21 & 20,40,11,31,51_2 & 
10,30,50,01,21_3,41 & 00,20,11_2,31,51
\end{array}
\label{TableEAxAE=EE}
$$
}

The reader who does not want to use the previous simple matrix 
manipulations to generate the whole table
may proceed as follows (this is equivalent):
First we remember that $S$ is both a left and right $E_6$ module and 
that $\sigma_0$ is the unit of $E_6$.
Therefore we have
 $$\sigma_a \odot \sigma_b = \sigma_a ( \sigma_0 \odot 
\sigma_0)\sigma_b =  \sigma_a ( 0\otimesdot 0 + 1 \otimesdot 1 
)\sigma_b$$
Let us compute for instance $\sigma_1 \odot \sigma_2$ and write only 
the subscripts to save space.
\begin{eqnarray*}
1\odot 2 &=& 1 ( 0 \otimesdot 0 + 1 \otimesdot 1) 2 = (1.0) 
\otimesdot (0.2) + (1.1) \otimesdot (1.2) \\
{} &=& 1  \otimesdot 2 + (0 + 2)  \otimesdot (1 + 3 + 5) \\
{} & = & 1  \otimesdot 2 + 0  \otimesdot 1 + 0  \otimesdot 3 + 0  
\otimesdot 5 + 2  \otimesdot 1 + 2  \otimesdot 3 +
 2  \otimesdot 5\\
{} & = & 2  \otimesdot 1 + 0  \otimesdot 1 + 3  \otimesdot 0 + 4  
\otimesdot 1 + 2  \otimesdot 1 + 1  \otimesdot 0 
+ 5  \otimesdot 0 + 2  \otimesdot 1 
\end{eqnarray*}
At the last line, we used the fact that, in $S$, $1 \otimesdot 2 = 2 
\otimesdot 1$,
 $0 \otimesdot 3 = 3 \otimesdot 0$,  $0 \otimesdot 5 = 4 \otimesdot 
1$,
 $2 \otimesdot 3 = 2.3  \otimesdot 0 = 1  \otimesdot 0 + 5  
\otimesdot 0 $ and that
$2  \otimesdot 5 = 2  \otimesdot 4.1 = 2.4  \otimesdot 1 = 2  
\otimesdot 1$.

Notice that this table looks very much like the table that was called 
$E_a \times \widetilde E_b$ (or $E_6 \times E_6 \mapsto A_{11}$)
 in a previous section, but
now, lengths of essential paths are replaced by the $x = a \otimesdot b$ 
labels of the algebra $S$.
We could also encode this structure in terms of six matrices of 
dimension $(12,6)$ (exactly as we encoded all data concerning
essential paths in terms of six matrices of dimension $(11,6)$), now 
the rows would be labelled by the twelve basis elements of $S$ rather than by 
the essential paths of $A_{11}$; in other words, the twelve matrices 
$S_{x = a \otimesdot b}$ replace the eleven matrices 
$F^{E_{6}}_{n}$.

We can also read the previous table in terms of essential  matrices and 
graph fusion matrices:
Multiplying the transpose of a matrix $E_a$ (of dimension $(6,11)$) 
by  a matrix $E_b$ (of dimension $(11,6)$)
gives a matrix $(6,6)$ and it can indeed be checked that the above 
table gives the decomposition of a product
$\widetilde E_a^{E_6}. E_b^{E_6}$ in terms of (product of) graph fusion matrices 
$N_c^{E_{6}}$ for the graph $E_6$.
For instance 
$$\widetilde E_1^{E_6}  E_5^{E_6} = N_2^{E_{6}} N_0^{E_{6}} +  
N_4^{E_{6}} N_0^{E_{6}} +
 N_1^{E_{6}} N_1^{E_{6}} + N_3^{E_{6}} N_1^{E_{6}} + 2 N_{5}^{E_{6}} 
N_1^{E_{6}}$$
For this reason the above table  $E_6 \times E_6 \mapsto S$ may also 
be called ``the $\widetilde E_a \times E_b$ table''.
More generally, these relations read:
$$
\widetilde E_a^{E_6}  E_b^{E_6} = \sum_{x \in Oc(E_{6})}
(S_{x})_{ab} \, S_{x}
$$
Here the sum is over all twelve vertices $x$ of the Ocneanu graph of 
$E_{6}$ (compare with section \ref{xiab}).
Once the $E_{6} \times E_{6} \rightarrow S$ table is known (and we 
have explained how to get it), it is a simple matter to check
all these equations for $\widetilde E_a^{E_6}  E_b^{E_6}$.
Admitedly, this is not an enlighting method. The simplest and
most direct proof uses the fact that $E_{a}^{E_{6}} = E_{0}^{E_{6}} . 
N_{a}^{E_{6}}$ and that (only one equation to check) $$
\widetilde E_0^{E_6}  E_0^{E_6} = S_{0\otimesdot 0} + S_{1\otimesdot 
1} = N_{0}^{E_6} + N_{0}^{E_6} + N_{2}^{E_6}
$$
The conclusion follows from the fact that $S_{a\otimesdot b}=N_{a}^{E_6}.N_{b}^{E_6}$.
A more formal proof (which would not use the above explicit result for $\widetilde 
E_0^{E_6}  E_0^{E_6}$) cannot be as straightforward as 
its analog (for $E_a^{E_6} \widetilde E_b^{E_6}$) presented in section 
\ref{xiab}; the difficulty, now, is to relate the eleven matrices $F_{n}^{E_{6}}$ to the twelve 
matrices $S_{x}$ (this can be a posteriori done in terms of explicit 
Fourier-like matrices $11\times 12$). In any case, the proof using the 
explicit calculation of $\widetilde E_0^{E_6}  E_0^{E_6}$ is easy.

Here again it is handy to describe the non zero entries of the  table 
by a new kind of vertices.
This should be compared with those introduced in section \ref{xiab}.
The former table ($E_6 \times E_6 \mapsto A_{11}$) gives all possible 
vertices of the type displayed in
sect. 3.4, where they are associated with nonzero matrix elements 
$(F_{n}^{E_{6}})_{ab}$; entries of
the later table ( $E_6 \times E_6 \mapsto S$) are 
associated with nonzero matrix elements 
$(S_{x}^{E_{6}})_{ab}$ and gives all 
possible vertices of the following type \label{fig:EES}:

\includegraphics{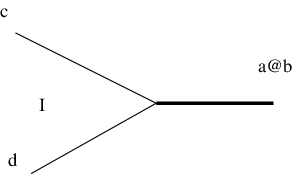}
\smallskip
We introduce  one such vertex,
  called  $I_{cd}^{a\otimesdot b}$,
 whenever $a\stackrel{\cdot}{\otimes} b$ appears 
in the previous $\widetilde E_c \times E_d$-table, at the intersection of 
row $c$ and column $d$.
We have therefore $6$ vertices of type $I_{\cdot \cdot}^{0 \otimesdot 
0}$, $8$ vertices of type  $I_{\cdot \cdot}^{3 \otimesdot 0}$,
\etc
 The integer $d_{a\otimesdot b}$ is also the number of vertices of 
type  $I_{\cdot \cdot}^{a \otimesdot b}$; we find

$$
\left( \begin{array}{c|cccccccccccc}
   a\otimesdot b &  0 \otimesdot 0 &     3 \otimesdot 0 &     4 
\otimesdot 0 &     1 \otimesdot 0 &     2 \otimesdot 0 &     5 
\otimesdot 0 &     0 \otimesdot 1 &     0 \otimesdot 2 &     0 
\otimesdot 5 &     1 \otimesdot 1 &     2 \otimesdot 1 &     5 
\otimesdot 1 \\
   d_{a \otimesdot b} &  6 & 8 & 6 & 10 & 14& 10 & 10 & 14 & 10 & 20  
& 28 & 20
\end{array} \right)
$$

The sum of the squares of these numbers is $\sum_{a\otimesdot b \in 
S} d_{a\otimesdot b}^2 = 2512 $. 

We notice immediately that this sum is also equal to the sum  
$\sum_{n\in A_{11}} d_{n}^2 = 2512 $ found previously
in the section devoted to the study of essential paths.

\subsection{The algebra ${\cal A}$ of endomorphisms of essential 
paths (again)}

The fact that  $\sum_{a\otimesdot b \in S} d_{a\otimesdot b}^2 =  
\sum_{n\in A_{11}} d_{n}^2  (= 2512) $ suggests immediately
that the algebra  ${\cal A} = \bigoplus_n EssPath^n \otimes 
EssPath^n$ carries {\sl two}  algebra structures :
For the first stucture, it is a direct sum of eleven blocks  (square 
matrices) of dimensions $d_n$; for the second structure, 
 it is a direct sum of twelve blocks  (square matrices) of dimensions 
$d_{a\otimesdot b}$. In other words,
rather than decomposing  ${\cal A}$ on the tensor products
$\xi \otimes \eta$, as we did in section \ref{xiab}, we  decompose it 
on tensor products $I_1 \otimes I_2$, where $I_1$ and $I_2$ 
refer to the vertices appearing in the $E_6 \times E_6 \mapsto S$ 
table.
The two decompositions of the  {\sl bi-algebra} ${\cal A}$ as two 
distinct sums of blocks 
correspond to a diagonalisation of the two algebra structures.
The first algebra structure is the composition of endomorphisms of 
essential paths, 
and it is directly given by the very definition of ${\cal A}$;
its block decomposition is labeled by the points of $A_{11}$.
The second algebra structure (call it $\ast$) comes from the fact 
that essential paths (on which the elements of ${\cal A}$
act) are endowed with a partial multiplication, namely concatenation 
of paths, and one can use this to define, by duality,
the new multiplication on ${\cal A}$. One technical difficulty is 
that the concatenation of two
essential paths is not necessarily essential, so that one has to 
reproject the result of concatenation to obtain an essential
path.  We shall not describe explicitly this construction and refer 
to \cite{Ocneanu:paths}.
The block decomposition of ${\cal A}$, with respect to the second 
algebra
structure is labelled by the basis elements $J=a\otimesdot b$  of 
$S$:  
${\cal A} = \bigoplus_J {\cal A}_J = \bigoplus_J H_J \otimes H_J$. 
The index $J$ labelling the different blocks
is therefore also associated with minimal central projectors for the 
product $\ast$ (for instance, 
the central projector associated with the
block of dimension $28^2$ is a direct sum of twelve matrices, its 
restriction to this chosen block is the identity matrix, and
all other blocks are zero).
This is (probably) how the Ocneanu graph of quantum symmetries was 
first defined and obtained. 
From a pictural point of view, the dimension of the vector space
$H_{a\otimesdot b}$ is $d_{a\otimesdot b}$ and given, as we know, by 
the number of all vertices of type $I_{cd}^{a\otimesdot b}$;
The dimension of the block ${\cal A}_{a\otimes b} =H_{a\otimesdot 
b}\otimes H_{a\otimesdot b} $ is therefore given by
 the counting of all the possible labelled dual diffusion graphs with 
fixed internal line labelled $a\otimesdot b$.
Elements of  ${\cal A}_{a\otimes b}$  can be depicted by the 
following figure that looks like a {\sl dual} diffusion
graph of  particle physics.
 
\includegraphics{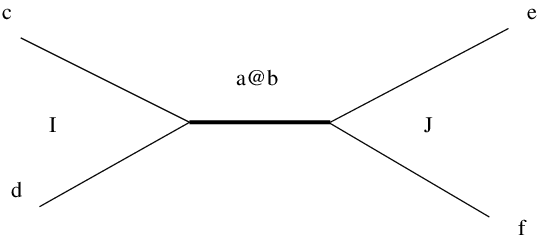}

\section{Modular structure of the $E_6$ graph}
\subsection{Verlinde representation}
\label{sec:Verlinde}

Consider the following $11\times 11$ matrices $S$ and $T$ (here $N = 
12$)
\begin{eqnarray*}
S[m,n ]&=& (-2I/(\sqrt 2   \sqrt N   ) )Sin[\pi \, m \, n/N] \\
T[m ,n ]&=& \exp[I \pi (m^2/(2 N )+ 1/4)] \delta_{m,n}
\end{eqnarray*}

We may check that $S^4=1$ and that $(ST)^3 = 1$;
 we have also $S^2 = -1$. The above matrices $S$ and $T$ 
are therefore  representatives for the generators
$\pmatrix 0 & 1 \\ -1 & 0 \endpmatrix$ and $\pmatrix 1 & 1 \\ 0 & 1 
\endpmatrix$ 
 of the modular group.
 This is an eleven dimensional representation of the group 
$SL(2,\ZZ)$.

The above $11\times 11$ matrix $S$ is nothing else than the ``non 
commutative character
table'' of the graph $A_{11}$: one may indeed check that the eleven 
matrices $S^{-1}.N_n^{A_{11}}.S$
are diagonal (like in \ref{sec:A11},  the $N_n^{A_{11}}$ denote the 
fusion matrices for the graph $A_{11}$).

As for the diagonal $T$ matrix, its eigenvalues can be
directly obtained from the central element
that defines the ribbon structure of the quantum group $U_q(SL2)$.
One may notice that eigenvalues of $T$ corresponding to 
irreducible
representations of $A_{11}$ associated by induction (see figure
 \ref{fig:EssPath0}) 
with the three extremal
points of the graph $E_6$ (the $A_3$ subalgebra) are equal. This is 
not
so for the other points of the $E_6$ graphs.

One may also notice that in the above representation of the modular 
group,  $T^{48}=1$ (and $T^s \neq 1$ for  smaller powers of $T$); 
one can actually prove\cite{Hurwitz} that the representation factors over the finite quotient 
$SL(2,\ZZ/48 \ZZ)$.

\subsection{Reduced essential matrices}

The reduced essential matrices, in the case of $E_6$, are obtained 
(definition)
from the essential matrices by 
putting to zero all the matrix elements of
 the columns associated with vertices 
$\{\sigma_1,\sigma_2,\sigma_5\}$.
Because of the order chosen for basis elements, this corresponds 
to columns $2,3,4$ of our essential matrices.
In general the reduced essential matrices $E^{r}_a$ should be obtained by
keeping only the matrix elements of the columns relative to the 
subalgebra defining the ambichiral part of the Ocneanu graph, and 
putting all others to zero.
In our example,  this subalgebra is $A_3$ and spanned by 
$\sigma_0,\sigma_4,
\sigma_3$. 
We shall see later why it is useful to  introduce these objects.

{\small

$$
\begin{array}{ccc}
E^{r}_0 = 
\left( \begin{array}{cccccc}
   1 & . & . & . & . & . \cr . & 
   . & . & . & . & . \cr . & . & 
   . & . & . & . \cr . & . & . & 
   . & . & 1 \cr . & . & . & . & 
   1 & . \cr . & . & . & . & . & 
   . \cr 1 & . & . & . & . & . \cr
   . & . & . & . & . & 1 \cr . & 
   . & . & . & . & . \cr . & . & 
   . & . & . & . \cr . & . & . & 
   . & 1 & . \cr
 \end{array} \right)

&

E^{r}_1 = 
\left( \begin{array}{cccccc}
 . & . & . & . & . & . \cr 1 & 
   . & . & . & . & . \cr . & . & 
   . & . & . & 1 \cr . & . & . & 
   . & 1 & . \cr . & . & . & . & 
   . & 1 \cr 1 & . & . & . & 1 & 
   . \cr . & . & . & . & . & 1 \cr
   1 & . & . & . & . & . \cr . & 
   . & . & . & . & 1 \cr . & . & 
   . & . & 1 & . \cr . & . & . & 
   . & . & . \cr   
 \end{array} \right)

&

E^{r}_2 = 
\left( \begin{array}{cccccc}
    . & . & . & . & . & . \cr . & 
   . & . & . & . & 1 \cr 1 & . & 
   . & . & 1 & . \cr . & . & . & 
   . & . & 1 \cr 1 & . & . & . & 
   1 & . \cr . & . & . & . & . & 
   2 \cr 1 & . & . & . & 1 & . \cr
   . & . & . & . & . & 1 \cr 1 & 
   . & . & . & 1 & . \cr . & . & 
   . & . & . & 1 \cr . & . & . & 
   . & . & . \cr 
 \end{array} \right)

\\
{} & {} & {} 
\\
E^{r}_5 = 
\left( \begin{array}{cccccc}
  . & . & . & . & . & . \cr . & 
   . & . & . & 1 & . \cr . & . & 
   . & . & . & 1 \cr 1 & . & . & 
   . & . & . \cr . & . & . & . & 
   . & 1 \cr 1 & . & . & . & 1 & 
   . \cr . & . & . & . & . & 1 \cr
   . & . & . & . & 1 & . \cr . & 
   . & . & . & . & 1 \cr 1 & . & 
   . & . & . & . \cr . & . & . & 
   . & . & . \cr
 \end{array} \right)

&

E^{r}_4 = 
\left( \begin{array}{cccccc}
. & . & . & . & 1 & . \cr . & 
   . & . & . & . & . \cr . & . & 
   . & . & . & . \cr . & . & . & 
   . & . & 1 \cr 1 & . & . & . & 
   . & . \cr . & . & . & . & . & 
   . \cr . & . & . & . & 1 & . \cr
   . & . & . & . & . & 1 \cr . & 
   . & . & . & . & . \cr . & . & 
   . & . & . & . \cr 1 & . & . & 
   . & . & . \cr
 \end{array} \right)

&

E^{r}_3 = 
\left( \begin{array}{cccccc}
. & . & . & . & . & 1 \cr . & 
   . & . & . & . & . \cr . & . & 
   . & . & . & . \cr 1 & . & . & 
   . & 1 & . \cr . & . & . & . & 
   . & 1 \cr . & . & . & . & . & 
   . \cr . & . & . & . & . & 1 \cr
   1 & . & . & . & 1 & . \cr . & 
   . & . & . & . & . \cr . & . & 
   . & . & . & . \cr . & . & . & 
   . & . & 1 \cr 
 \end{array} \right)

\end{array}
$$
}

\subsection{Torus structure of the $E_6$ graph}

A set of matrices describing what is called
``the torus structure of a Dynkin diagram'' 
was mentioned by A. Ocneanu in various talks since 1995 
(for instance \cite{Ocneanu:Marseille}),
but  --- to our knowledge --- this has not been made available in written 
form.
We shall therefore neither comment about the original definition of 
these matrices nor relate Ocneanu's construction to 
ours... but it is clear that the Ocneanu matrices describing the 
``torus structure'' of the Dynkin diagrams and our ``toric
matrices'' are the same objects.
We shall  introduce them directly,
in terms of our essential matrices and reduced essential matrices.

 To every point $a\otimesdot b$ of the Ocneanu graph of quantum 
symmetries  (\ie the twelve points corresponding to linear
generators of $S = E_6 \otimes_{A_3} E_6$, 
in the case of the Dynkin diagram $E_6$),
we associate a matrix $11\times 11$ (more generally a square matrix 
$(N-1) \times (N-1)$ if $N$ is the dual Coxeter number of the
chosen Dynkin diagram) defined by $W_{ab} \doteq E_a \widetilde E^{r}_b$ ( $ 
= E^{r}_a \widetilde E^{r}_b$ ). Starting from the point $a\otimesdot b$ 
of the Ocneanu graph, the number $(W_{ab})_{ij}$ counts the number of 
independent ways to reach the origin  $0\otimesdot 0$ after having 
performed essential paths of length $i$ (resp. $j$) on the left and 
right chiral subgraphs.

{\small
$$
\begin{array}{cc}

W_{00} & W_{11}  \\
{} & {} \\


\left( \begin{array}{ccccccccccc}
   1 & . & . & . & . & . & 1 & . &
   . & . & . \cr . & . & . & . & 
   . & . & . & . & . & . & . \cr 
   . & . & . & . & . & . & . & . &
   . & . & . \cr . & . & . & 1 & 
   . & . & . & 1 & . & . & . \cr 
   . & . & . & . & 1 & . & . & . &
   . & . & 1 \cr . & . & . & . & 
   . & . & . & . & . & . & . \cr 
   1 & . & . & . & . & . & 1 & . &
   . & . & . \cr . & . & . & 1 & 
   . & . & . & 1 & . & . & . \cr 
   . & . & . & . & . & . & . & . &
   . & . & . \cr . & . & . & . & 
   . & . & . & . & . & . & . \cr 
   . & . & . & . & 1 & . & . & . &
   . & . & 1 \cr  
 \end{array} \right)

&

\left( \begin{array}{ccccccccccc}
 . & . & . & . & . & . & . & . &
   . & . & . \cr . & 1 & . & . & 
   . & 1 & . & 1 & . & . & . \cr 
   . & . & 1 & . & 1 & . & 1 & . &
   1 & . & . \cr . & . & . & 1 & 
   . & 1 & . & . & . & 1 & . \cr 
   . & . & 1 & . & 1 & . & 1 & . &
   1 & . & . \cr . & 1 & . & 1 & 
   . & 2 & . & 1 & . & 1 & . \cr 
   . & . & 1 & . & 1 & . & 1 & . &
   1 & . & . \cr . & 1 & . & . & 
   . & 1 & . & 1 & . & . & . \cr 
   . & . & 1 & . & 1 & . & 1 & . &
   1 & . & . \cr . & . & . & 1 & 
   . & 1 & . & . & . & 1 & . \cr 
   . & . & . & . & . & . & . & . &
   . & . & . \cr 
 \end{array} \right)

\\

{} & {} \\
{} & {} \\
W_{30} & W_{21}  \\
{} & {} \\

 \left( \begin{array}{ccccccccccc}
   . & . & . & 1 & . & . & . & 1 &
   . & . & . \cr . & . & . & . & 
   . & . & . & . & . & . & . \cr 
   . & . & . & . & . & . & . & . &
   . & . & . \cr 1 & . & . & . & 
   1 & . & 1 & . & . & . & 1 \cr 
   . & . & . & 1 & . & . & . & 1 &
   . & . & . \cr . & . & . & . & 
   . & . & . & . & . & . & . \cr 
   . & . & . & 1 & . & . & . & 1 &
   . & . & . \cr 1 & . & . & . & 
   1 & . & 1 & . & . & . & 1 \cr 
   . & . & . & . & . & . & . & . &
   . & . & . \cr . & . & . & . & 
   . & . & . & . & . & . & . \cr 
   . & . & . & 1 & . & . & . & 1 &
   . & . & . \cr
 \end{array} \right)

&

\left( \begin{array}{ccccccccccc}
 . & . & . & . & . & . & . & . &
   . & . & . \cr . & . & 1 & . & 
   1 & . & 1 & . & 1 & . & . \cr 
   . & 1 & . & 1 & . & 2 & . & 1 &
   . & 1 & . \cr . & . & 1 & . & 
   1 & . & 1 & . & 1 & . & . \cr 
   . & 1 & . & 1 & . & 2 & . & 1 &
   . & 1 & . \cr . & . & 2 & . & 
   2 & . & 2 & . & 2 & . & . \cr 
   . & 1 & . & 1 & . & 2 & . & 1 &
   . & 1 & . \cr . & . & 1 & . & 
   1 & . & 1 & . & 1 & . & . \cr 
   . & 1 & . & 1 & . & 2 & . & 1 &
   . & 1 & . \cr . & . & 1 & . & 
   1 & . & 1 & . & 1 & . & . \cr 
   . & . & . & . & . & . & . & . &
   . & . & . \cr 
 \end{array} \right)

\\
{} & {} \\
{} & {} \\
W_{40} & W_{51}  \\
{} & {} \\

\left( \begin{array}{ccccccccccc}
   . & . & . & . & 1 & . & . & . &
   . & . & 1 \cr . & . & . & . & 
   . & . & . & . & . & . & . \cr 
   . & . & . & . & . & . & . & . &
   . & . & . \cr . & . & . & 1 & 
   . & . & . & 1 & . & . & . \cr 
   1 & . & . & . & . & . & 1 & . &
   . & . & . \cr . & . & . & . & 
   . & . & . & . & . & . & . \cr 
   . & . & . & . & 1 & . & . & . &
   . & . & 1 \cr . & . & . & 1 & 
   . & . & . & 1 & . & . & . \cr 
   . & . & . & . & . & . & . & . &
   . & . & . \cr . & . & . & . & 
   . & . & . & . & . & . & . \cr 
   1 & . & . & . & . & . & 1 & . &
   . & . & . \cr
 \end{array} \right)

&

\left( \begin{array}{ccccccccccc}
   . & . & . & . & . & . & . & . &
   . & . & . \cr . & . & . & 1 & 
   . & 1 & . & . & . & 1 & . \cr 
   . & . & 1 & . & 1 & . & 1 & . &
   1 & . & . \cr . & 1 & . & . & 
   . & 1 & . & 1 & . & . & . \cr 
   . & . & 1 & . & 1 & . & 1 & . &
   1 & . & . \cr . & 1 & . & 1 & 
   . & 2 & . & 1 & . & 1 & . \cr 
   . & . & 1 & . & 1 & . & 1 & . &
   1 & . & . \cr . & . & . & 1 & 
   . & 1 & . & . & . & 1 & . \cr 
   . & . & 1 & . & 1 & . & 1 & . &
   1 & . & . \cr . & 1 & . & . & 
   . & 1 & . & 1 & . & . & . \cr 
   . & . & . & . & . & . & . & . &
   . & . & . \cr 
 \end{array} \right)

\end{array}
$$
}

{\small

$$
\begin{array}{cc}

W_{10} & W_{01}  \\
{} & {} \\

\left( \begin{array}{ccccccccccc}
   . & . & . & . & . & . & . & . &
   . & . & . \cr 1 & . & . & . & 
   . & . & 1 & . & . & . & . \cr 
   . & . & . & 1 & . & . & . & 1 &
   . & . & . \cr . & . & . & . & 
   1 & . & . & . & . & . & 1 \cr 
   . & . & . & 1 & . & . & . & 1 &
   . & . & . \cr 1 & . & . & . & 
   1 & . & 1 & . & . & . & 1 \cr 
   . & . & . & 1 & . & . & . & 1 &
   . & . & . \cr 1 & . & . & . & 
   . & . & 1 & . & . & . & . \cr 
   . & . & . & 1 & . & . & . & 1 &
   . & . & . \cr . & . & . & . & 
   1 & . & . & . & . & . & 1 \cr 
   . & . & . & . & . & . & . & . &
   . & . & . \cr  
 \end{array} \right)

&

\left( \begin{array}{ccccccccccc}
 . & 1 & . & . & . & 1 & . & 1 &
   . & . & . \cr . & . & . & . & 
   . & . & . & . & . & . & . \cr 
   . & . & . & . & . & . & . & . &
   . & . & . \cr . & . & 1 & . & 
   1 & . & 1 & . & 1 & . & . \cr 
   . & . & . & 1 & . & 1 & . & . &
   . & 1 & . \cr . & . & . & . & 
   . & . & . & . & . & . & . \cr 
   . & 1 & . & . & . & 1 & . & 1 &
   . & . & . \cr . & . & 1 & . & 
   1 & . & 1 & . & 1 & . & . \cr 
   . & . & . & . & . & . & . & . &
   . & . & . \cr . & . & . & . & 
   . & . & . & . & . & . & . \cr 
   . & . & . & 1 & . & 1 & . & . &
   . & 1 & . \cr
 \end{array} \right)

\\
{} & {} \\
{} & {} \\
W_{20} & W_{02} = W_{31}  \\
{} & {} \\

\left( \begin{array}{ccccccccccc}
  . & . & . & . & . & . & . & . &
   . & . & . \cr . & . & . & 1 & 
   . & . & . & 1 & . & . & . \cr 
   1 & . & . & . & 1 & . & 1 & . &
   . & . & 1 \cr . & . & . & 1 & 
   . & . & . & 1 & . & . & . \cr 
   1 & . & . & . & 1 & . & 1 & . &
   . & . & 1 \cr . & . & . & 2 & 
   . & . & . & 2 & . & . & . \cr 
   1 & . & . & . & 1 & . & 1 & . &
   . & . & 1 \cr . & . & . & 1 & 
   . & . & . & 1 & . & . & . \cr 
   1 & . & . & . & 1 & . & 1 & . &
   . & . & 1 \cr . & . & . & 1 & 
   . & . & . & 1 & . & . & . \cr 
   . & . & . & . & . & . & . & . &
   . & . & . \cr  
 \end{array} \right)

&

\left( \begin{array}{ccccccccccc}
 . & . & 1 & . & 1 & . & 1 & . &
   1 & . & . \cr . & . & . & . & 
   . & . & . & . & . & . & . \cr 
   . & . & . & . & . & . & . & . &
   . & . & . \cr . & 1 & . & 1 & 
   . & 2 & . & 1 & . & 1 & . \cr 
   . & . & 1 & . & 1 & . & 1 & . &
   1 & . & . \cr . & . & . & . & 
   . & . & . & . & . & . & . \cr 
   . & . & 1 & . & 1 & . & 1 & . &
   1 & . & . \cr . & 1 & . & 1 & 
   . & 2 & . & 1 & . & 1 & . \cr 
   . & . & . & . & . & . & . & . &
   . & . & . \cr . & . & . & . & 
   . & . & . & . & . & . & . \cr 
   . & . & 1 & . & 1 & . & 1 & . &
   1 & . & . \cr 
 \end{array} \right)

\\

{} & {} \\
{} & {} \\
W_{50} & W_{05} = W_{41}  \\
{} & {} \\

\left( \begin{array}{ccccccccccc}
   . & . & . & . & . & . & . & . &
   . & . & . \cr . & . & . & . & 
   1 & . & . & . & . & . & 1 \cr 
   . & . & . & 1 & . & . & . & 1 &
   . & . & . \cr 1 & . & . & . & 
   . & . & 1 & . & . & . & . \cr 
   . & . & . & 1 & . & . & . & 1 &
   . & . & . \cr 1 & . & . & . & 
   1 & . & 1 & . & . & . & 1 \cr 
   . & . & . & 1 & . & . & . & 1 &
   . & . & . \cr . & . & . & . & 
   1 & . & . & . & . & . & 1 \cr 
   . & . & . & 1 & . & . & . & 1 &
   . & . & . \cr 1 & . & . & . & 
   . & . & 1 & . & . & . & . \cr 
   . & . & . & . & . & . & . & . &
   . & . & . \cr 
 \end{array} \right)

&

\left( \begin{array}{ccccccccccc}
   . & . & . & 1 & . & 1 & . & . &
   . & 1 & . \cr . & . & . & . & 
   . & . & . & . & . & . & . \cr 
   . & . & . & . & . & . & . & . &
   . & . & . \cr . & . & 1 & . & 
   1 & . & 1 & . & 1 & . & . \cr 
   . & 1 & . & . & . & 1 & . & 1 &
   . & . & . \cr . & . & . & . & 
   . & . & . & . & . & . & . \cr 
   . & . & . & 1 & . & 1 & . & . &
   . & 1 & . \cr . & . & 1 & . & 
   1 & . & 1 & . & 1 & . & . \cr 
   . & . & . & . & . & . & . & . &
   . & . & . \cr . & . & . & . & 
   . & . & . & . & . & . & . \cr 
   . & 1 & . & . & . & 1 & . & 1 &
   . & . & . \cr 
 \end{array} \right)

\end{array}
$$
}

\subsection{The modular invariant}

The toric matrix associated with the origin $0 \otimesdot 0$ of the 
Ocneanu graph of $E_6$
is
$$
W_{00} =
\left( \begin{array}{ccccccccccc}
   1 & . & . & . & . & . & 1 & . &
   . & . & . \cr . & . & . & . & 
   . & . & . & . & . & . & . \cr 
   . & . & . & . & . & . & . & . &
   . & . & . \cr . & . & . & 1 & 
   . & . & . & 1 & . & . & . \cr 
   . & . & . & . & 1 & . & . & . &
   . & . & 1 \cr . & . & . & . & 
   . & . & . & . & . & . & . \cr 
   1 & . & . & . & . & . & 1 & . &
   . & . & . \cr . & . & . & 1 & 
   . & . & . & 1 & . & . & . \cr 
   . & . & . & . & . & . & . & . &
   . & . & . \cr . & . & . & . & 
   . & . & . & . & . & . & . \cr 
   . & . & . & . & 1 & . & . & . &
   . & . & 1 \cr  
 \end{array} \right)
$$

It can be checked that it commutes with $S$ and $T$, and therefore 
with the whole $SL_2(\ZZ)$ group.
\begin{eqnarray*}
S \, W_{00} &=& W_{00} \, S \\
T \, W_{00} &=& W_{00} \, T \\
\end{eqnarray*}
Notice that  $ W_{00}$ is  normalized ($W_{00}[1,1]=1$)
and that all the entries of this matrix are {\sl positive} integers.
The following $SL(2,\ZZ)$-invariant sesquilinear form on $\CC^{11}$
$$
Z = \sum_{i,j=1}^{11} W_{00}[i,j] \, \chi^i \overline \chi^j = \vert 
\chi_1 + \chi_7 \vert^2 +
 \vert \chi_4 + \chi_8 \vert^2 +
 \vert \chi_5 + \chi_{11} \vert^2
$$
gives a solution of  the Cappelli-Itzykson-Zuber problem and
 can therefore be interpreted as the modular invariant partition 
function of a quantum field model.

This partition function was of course obtained long ago but it is 
interesting to notice that it can be recovered
as the toric matrix associated with the origin of the 
Ocneanu graph of $E_6$.

As stated in \cite{Ocneanu:paths}, each entry of  $W_{00}[i,j]$ can 
be considered as the dimension of a representation of
the algebra of quantum symmetries (for instance, in the case of 
$E_6$, we have twelve non-zero components and all the components are
equal to $1$, this reflects the fact that $S$ is a twelve dimensional 
abelian algebra). Notice finally that the Coxeter numbers
of the graph ($1,4,5,7,8,11$) appear on the diagonal of this matrix.

The above reconstruction of the partition function also works for all 
ADE (\ie chiral algebra of $\hat SU(2)$ type) statistical models
 and can probably be generalized to more general situations ( $\hat 
SU(3)$, \ldots).
The interpretation of the other toric matrices $W_{ab}$, in terms of 
conformal field theory models, still requires some work \footnote{See 
however the last footnote of section \ref{sec:remarks}}.

\section{Miscellaneous comments}

\subsection{Remarks about conformal embeddings}

The condition for conformal embeddings of affine Kac Moody algebras 
${\cal G}_1 \subset {\cal G}_2$ at levels $k_1$ and $k_2$
is the identity $c_1 = c_2$ of their central charges, where $c$ is 
given by $c=\frac{k\, dim({\cal G})}{k + \hat g}$, 
$\hat g$ being the dual Coxeter number of the corresponding finite 
dimensional Lie algebra.
If one takes ${\cal G}_1 = \hat A_1$ and $k_2 = 1$, there are only 
three non trivial solutions (of course levels should be integers):
$(\hat A_1)_{10}  \subset (\hat B_2)_{1} = (\hat C_2)_{1}$,  $(\hat 
A_1)_{28}  \subset (\hat G_2)_{1}$ and
 $(\hat A_1)_{4}  \subset (\hat A_2)_{1}$,
 since Coxeter numbers (resp. dimensions)
 of $A_1$, $A_2$, $B_1$ and $G_2$ are given by $2,3,3,4$ (resp. 
$3,8,10,14$). 
The first two conformal embeddings are respectively described by
the Dynkin diagrams of $E_6$ and $E_8$; the last one gives rise to 
the $D_4$ invariant.
This was observed in the literature long ago ~\cite{Nahm}
 and discussed,
in terms of inclusions of Von Neumann algebras,
 by ~\cite{Evans}, see also ~\cite{Evans:Bariloche} and references 
therein.

\subsection{Remarks about generalized quantum recoupling theory}

\subsubsection{Wigner and Racah multiplications: the pure $SU(2)$ 
case}

Using all possible  spin triples  $j_{1},j_{2},J$ such that  
 $\vert j_{1}-j_{2} \vert \leq J \leq \vert j_{1}+j_{2} \vert$, 
we can draw elementary vertices decorated with such allowed triples and
then build elementary ``diffusion graphs'' looking
like the following one \label{DiffusionGraphSU2}. Spin values 
(integer or half-integers) describe, as 
usual, the irreducible representations of $SU(2)$.

\includegraphics*{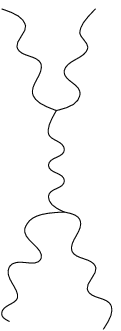}

Since we have 
infinitely many possible (labelled)
vertices at our disposal, we have also infinitely many such diffusion 
graphs. Since a triangle inequality has to be satisfied at each 
vertex, the above diffusion graphs could also be called (dually) ``double 
triangles'' and pictured as two triangles
glued together, sharing this time an horizontal common edge. 

Following Ocneanu we define a vector space
$\cal A$ generated by a linear basis indexed by
 the (infinite) set of diffusion graphs ---or double triangles. The 
vector space $\cal A$ comes therefore equiped with a particular 
basis, and every element of $\cal A$ is a linear combination, over 
the complex numbers, of diffusion graphs.
One also introduces another class of graphs: the ``dual'' diffusion 
graphs (here the internal line is horizontal, not vertical).

 The next step is to endow the vector space $A$ with {\sl two} 
compatible multiplications.
 These two multiplications appear in the works of Racah and Wigner 
and  are
mentionned in the book \cite{Biedenharn-Louck}.
The first multiplication (Wigner) amounts to compose these ``spin 
diffusion graphs'' vertically, 
the other (Racah) amounts to compose them horizontally. The 
precise definition involves appropriate coefficients
and we refer to the book ~\cite{Biedenharn-Louck} for explicit 
formulae. The following identity (a kind of Fourier transform)
relates the previous diffusion graphs and their dual 

\includegraphics*{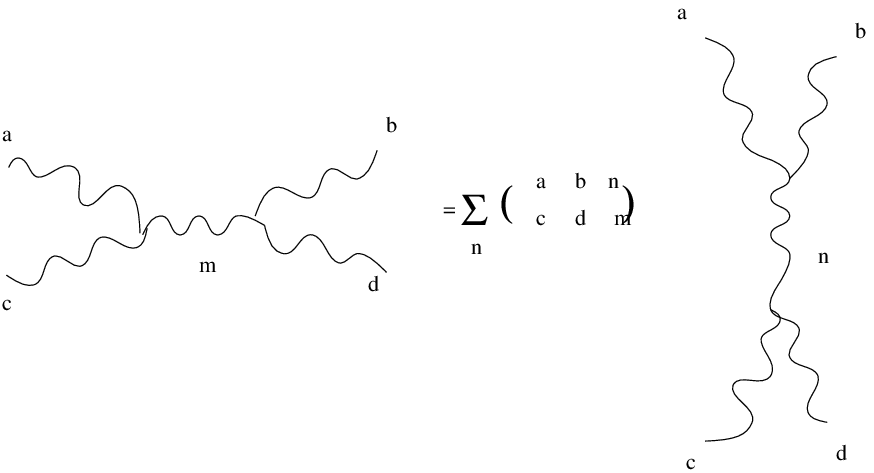}
\label{horiztovert}

Although the subject itself is quite old, 
 and besides a few lines lines in the book just quoted,
 even in the pure $SU(2)$ case
we cannot give any reference providing a precise study of
 this bigebra structure (that we suggest to call the ``Racah-Wigner 
bigebra'')

\subsubsection{From $SU(2)$ to its finite subgroups}

Rather than using representations of $G=SU(2)$, we can 
use the data provided by a finite 
subgroup $K$ of $SU(2)$ and consider several variants of the above 
bigebra.

\begin{itemize}
\item We may use irreps of $K$ alone.
\item We may use simultaneously irreps of $K$ and of $G$. Indeed, an irrep of $G$, 
restricted to $K$, can be tensorially multiplied by irreps of $K$ and
the result can be decomposed into irreps of $K$.
\item We may even choose {\sl two} subgroups $H$ and $K$ of $SU(2)$ (for instance 
the binary tetrahedral and binary icosahedral groups).
\end{itemize}

\subsubsection{Generalized quantum recoupling theory and Ocneanu 
bigebra}

The next step is to replace $G$ by the fusion algebra of the graph 
$A_N$ and $K$ by the fusion algebra of another graph with the
same norm (the norms have to match, otherwise, one cannot consider 
meaningful bimodules).
The ideas are the same as before,  but instead of the usual 
triangular condition of
	composition of spins (case of $SU(2)$), we have  more complicated
	conditions which are direct consequence of the multiplication tables 
(like those expressing the fusion rules
$A_{11} \times A_{11} \mapsto A_{11}$, $E_6 \times E_6 \mapsto E_6$ , 
$A_{11} \times E_6 \mapsto E_6$)
that we have considered previously). In the same way as for $SU(2)$, one 
may consider double triangles, generalized $6\,j$ symbols
(represented by tetrahedra carrying bimodules labels on their six  
edges, and algebra labels at the four vertices) \etc
The construction of the Ocneanu bigebra ${\cal A}$ is a direct 
generalization of the Racah-Wigner bigebra. 
It was introduced before in terms of endomorphisms of essential 
paths. 
Its dimension is finite ($2512$ in the case of $E_6$). The 
diagonalization of this algebra, for the product $\ast$, 
is ${\cal A}= \oplus_{J\in S} End(H_J)$. The Hilbert spaces $H_J$ are 
labelled by minimal central projections of $({\cal A},\ast)$.
Elements of ${\cal A}$, in particular the particular basis elements 
described by diffusion graphs (also seen as particular
endomorphisms of the space of essential paths or as two triangles
with vertices  $A_{11}, A_{11}, E_6$ sharing a common edge of type 
$A_{11} - A_{11}$), can be decomposed into linear combinations
of elements in $H_J \otimes H_J$, \ie in terms of ``dual diffusion 
graphs''.
This decomposition  generalizes the  equation given in 
\ref{horiztovert} and involves generalized $6\,j$ symbols 
(see ~\cite{Ocneanu:paths} for details).

\subsubsection{Connections on Ocneanu cells, quantum symmetries}
We  just give here the following references:
The general notion of connections on a system of four graphs was 
introduced in ~\cite{Ocneanu:paragroups}. 
Such ``cells'' have been explicitly written and used to reconstruct 
systems of Boltzmann weights in ~\cite{Pasquier}.
The general formalism of Ocneanu cells was ``translated''
and  adapted to the situation of statistical mechanics in 
~\cite{Roche}. In the 
framework of RSOS models, an explicit description of several cell systems 
can be found in \cite{Pearce-Zhou}.
The notion of quantum symmetries on such a system is due to A. 
Ocneanu; it was presented at several meetings since 1995 and it is 
described in the recent article  ~\cite{Ocneanu:paths}.

\subsection{Graph algebras and finite dimensional quantum groups}

The reader will have noticed that we did not provide, so far, any 
quantum group (Hopf algebra) interpretation for the above 
constructions.
 The main reason is that such a purely algebraic interpretation, in 
terms of finite dimensional --- but not necessarily
semi-simple --- Hopf algebras is not known for arbitrary ADE graphs. 
However, for $A_N$ Dynkin diagrams,  such an interpretation can be 
found. Here it is: Take the quantized enveloping algebra 
${\cal U} \doteq U_q(SL2)$ at a primitive $(N+1)$-th roots of unity.  
Take $N+1$ odd for the moment
 (the analysis can be done for even $N+1$, but this is slightly more
involved). Call $\cal H$ the quotient of $\cal U$ by the ideal 
generated by relations $K^{N+1} = 1$, $X_{\pm}^{N+1} = 0$ (here $K$ 
and $X_\pm$ denote the usual generators of $\cal U$). 
This ideal $\cal H$ 
is a Hopf ideal and the quotient is a finite dimensional Hopf 
algebra of
dimension $(N+1)^3$. As an algebra, it is isomorphic with $M(N+1) 
\oplus M(1\vert N)_0 \oplus M(2\vert N-1)_0 \oplus \ldots$ where
the first term is the algebra of $(N+1)\times (N+1)$ matrices over 
complex numbers and where $M(p\vert N+1-p)_0$ is the even part of the
algebra of $(N+1)\times (N+1)$ matrices with elements in the Grassman 
algebra with two
 generators (\cite{Alekseev},\cite{Ogievetsky:SL2}) \footnote{More 
information about $\cal H$ can be 
gathered from \cite{CGT:SL2}, \cite{CoqueSchieber:pol}  and references 
therein}.
This algebra is not semi-simple.
For instance, if $q^5=1$, ${\cal H} = 
M(5) \oplus M(4\vert 1)_0 \oplus M(3\vert 2)_0$. 
Projective indecomposable modules of this algebra are given by the 
columns, so (let us continue our example with $q^5$ = 1), 
we get one irreducible and projective representation of dimension $5$ 
and four inequivalent 
projective indecomposable representations of dimension $2\cdot 5 
=10$. Irreducible representations are obtained by taking the quotient
of each projective representation by its own radical; in this way we 
get finite dimensional irreducible (but not projective)
 representations of dimensions $1,2,3,4$. We label each projective 
indecomposable by the corresponding irreducible, so, besides
the particular $5$ dimensional representation, that is both 
irreducible and projective, we have four projective indecomposable
labelled  $10_1$, $10_2$,$10_3$,$10_4$. The notion of quantum 
dimension makes sense for this algebra; all the projective 
indecomposable representations (including the $5$-dimensional irrep) 
have quantum dimension $0$. The four irreducible representations of 
dimensions $1,2,3,4$ are of $q$-dimension $[1]_5, [2]_5, [3]_5, 
[4]_5$. We can tensorially multiply these representations and draw,
in particular, the diagram of tensorialisation (up to equivalence of 
representations) 
by  the $2$ dimensional irrep. Here is the diagram that we get:

\includegraphics{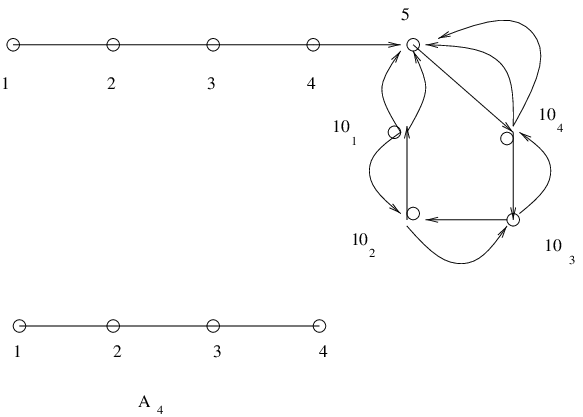}


If we now decide to discard the representations of $q$-dimension $0$, 
therefore 
removing all the projective indecomposable,
in particular disregarding as well the special irrep of dimension 
$5$, we just obtain the $A_4$ diagram. 
The fusion graph
of $A_4$ describes the tensor products of irreducible representations 
of ${\cal H}$ which are not of zero $q$-dimension.
In particular the equation $2 \otimes 2 \simeq  1 \oplus 3$ reads, in 
terms of $q$-dimensions (a quantity that is
multiplicative under tensor products): $\beta^2 = 1 + \beta$, and we 
recover the fact that the norm $\beta$ of $A_4$ is the
q-integer $[2]_5$, \ie  the golden number. The centralizer algebra, 
in the tensor powers (truncated as explained above)
of the fundamental representation of $\cal H$ is given by the Jones 
algebra for a particular value of the index ($1/\beta^2$).
Explicitly, this commutant is isomorphic with  $M(F_s,\CC) \oplus 
M(F_{s+1},\CC)$ where $F_s$ are Fibonacci numbers, since
$$
[2]^{2p} \simeq F_{2p-2} [1] + F_{2p-1} [3] , \quad [2]^{2p+1} \simeq 
F_{2p} [2] + F_{2p-1} [4]$$
Its dimension is no longer given by Catalan numbers (like for 
$SU(2)$), but by the sum of the squares
 of two consecutive Fibonacci number (so again a Fibonacci number).

Irreducible representations of $\cal H$ are particular irreducible 
representations $\rho$ of $\cal U$ 
(they are ``classical'' in the
sense that they are neither cyclic nor semi-cyclic). Moreover they 
are such that  $\omega \doteq \rho(K^{N+1}) = \pm 1.$
When $N+1$ is odd, in order to to get irreps with $\omega = -1$,
one has to replace the condition $K^{N+1}=1$ by $K^{2(N+1)}=1$ in the 
definition of ${\cal H}$ (we may call $\hat{\cal H}$ this
 algebra, whose dimension is twice the dimension of ${\cal H}$).
 When $N+1$ is even, the analysis is slightly different and they are 
two cases, depending upon the
parity of $(N+1)/2$. In any case, the tensor product of irreducible 
non projective representations of the finite dimensional 
Hopf algebra ${\cal H}$ can
be expressed in terms of the Dynkin diagram $A_N$ (with $q^{N+1} = 1$ 
if $N+1$ is odd, and $\qch^{2(N+1)}=1$ if $N+1$ is even).
In the case $q^{12} = \qch^{24} = 1$ the  algebra ${\cal H}$ is  
$M(1\vert 11)_0 \oplus M(3\vert 9)_0 \oplus \ldots \oplus 
M(9\vert 3)_0 \oplus M(11\vert 1)_0$, and $\hat{\cal H} ={\cal H} 
\oplus  M(0\vert 12) \oplus M(2\vert 10)_0
 \oplus \ldots M(10\vert 2)_0 \oplus M(12\vert 0)$.

In any case, the conclusion is that the fusion algebra of Dynkin 
diagrams of $A_N$ type can be given a purely finite dimensional
interpretation in terms of finite dimensional (not semi-simple) 
quantum groups. Such an interpretation is, at the moment, still
lacking in the case of $E_6$ or $E_8$, but we believe that it should 
be possible.
An interpretation of the algebraic constructions
of the type we considered in this paper can certainly be 
also formulated in
 categorical terms (truncated tensor products, braided categories 
\etc),  but we think that it is interesting to be able to use  finite 
dimensional Hopf algebras to describe such situations,
even if these Hopf algebras are not semi-simple. 
Notice that modules appearing in  discussions involving conformal 
embeddings (for instance $LSU(2)_{10} \subset LSpin(5)_1$)
are modules for affine algebras and are typically infinite 
dimensional.

\section{Conclusion}

The main results of this paper can be summarized as follows:
Take the Dynkin diagram $E_6$, consider its associated fusion algebra 
and its matrix realization (it is generated by
$G$, the adjacency matrix of the diagram);
call $E_a$ the essential matrices defined, for each vertex $a$ as the 
$11\times 6$ rectangular matrix
$E_a(row \, p) = E_a(row \, p-1).G - E_a(row \, p-2)$, where 
$E_a(row \, 0)$ is the row vector that lables the
chosen vertex $a$. We recover the Ocneanu graph of quantum symmetries 
of this Dynkin diagram as the Cayley graph
of multiplication by the two generators of the $12$-dimensional 
algebra $S=E_6 \otimes_{A_3} E_6$. The twelve
toric $11\times 11$ matrices $W_{a\otimesdot b}$ associated to the 
points of the Ocneanu graph can be obtained as $E_a.\widetilde E^{r}_b$
where the reduced essential matrices $E^{r}_b$ are gotten from the 
$E_a$'s by keeping only the columns associated
with the fusion subalgebra $A_3$. The toric matrix $W_{0\otimesdot 
0}$ is the modular invariant of $E_6$.
The choice of the $E_6$ example exhibits rather generic features, 
and it is not too hard to generalize the various
constructions to other cases (see the subsequent paper 
\cite{Coque-Gil:ADE})\footnote{The present paper was posted to the   
archives in November 2000 and several articles mentionned in the coming 
references section were not available at 
that time; for the convenience of the 
reader, we have added these papers to our references when the final 
version of our paper was sent to the publisher}.

\section{Acknowledgments}

This set of notes was elaborated during my stay at CERN 
and at the Mathematics Department
of the University of Geneva during the 
academic year 1999/2000. I would like to thank both places for 
their hospitality.
The author acknowledges support from CERN and from the Swiss National 
Science Foundation.
This research was also partly supported by a PICS contract No $608$, 
with Russia (Dubna).

I want to thank A. Ocneanu for his generous 
explanations  during his stay at C.P.T.  in Marseille (1995) and one 
of the referees for his constructive remarks.

\vfill
\eject

\end{document}